%% file: ms-v2.tex
\documentclass[iop]{emulateapj}

\PassOptionsToClass{10pt,aps,pra,floatfix}{revtex4-1} 

\usepackage{graphicx}
\usepackage[sumlimits,nointlimits,namelimits]{amsmath}

\usepackage{natbib}
\input{useful_macros}

\def\RLREF{REF\xspace}	
\def\AIREF{AM13-REF\xspace}

\pdfoutput=1

\begin{document}
\title{A Common Origin for Globular Clusters and Ultra-faint Dwarfs in  Simulations of the First Galaxies}
\shorttitle{Globular Clusters and Ultra-faint Dwarfs in Simulations of the First Galaxies} \shortauthors{Ricotti, Parry \& Gnedin}

\author{Massimo Ricotti \altaffilmark{1}, Owen H. Parry \altaffilmark{1}, Nickolay Y. Gnedin\altaffilmark{2,3,4}}
\affiliation{}
\altaffiltext{1}{Department of Astronomy, University of Maryland, College Park, MD 20740, USA; \mbox{\tt ricotti@astro.umd.edu}, \mbox{oparry@umd.edu}}
\altaffiltext{2}{Particle Astrophysics Center, Fermi National Accelerator Laboratory, Batavia, IL 60510, USA}
\altaffiltext{3}{Kavli Institute for Cosmological Physics and Enrico Fermi Institute, The University of Chicago, Chicago, IL 60637 USA}
\altaffiltext{4}{Department of Astronomy \& Astrophysics, University of Chicago, Chicago, IL 60637 USA}


\begin{abstract}
In this paper, the first in a series on galaxy formation before
reionization, we focus on understanding what determines the size and
morphology of stellar objects in the first low mass galaxies, using
parsec-scale cosmological simulations performed with an adaptive mesh
hydrodynamics code.  Although the dense gas in which stars are formed
tends to have a disk structure, stars are found in spheroids with
little rotation. Halos with masses between \appmass{6} and
\solmass{5}{8} form stars stochastically, with stellar masses in the
range \appmass{4} to \solmass{2}{6}. Nearly independent of stellar
mass, we observe a large range of half-light radii for the stars, from
a few parsecs to a few hundred parsecs and surface brightnesses and
mass-to-light ratios ranging from those typical of globular clusters
to ultra-faint dwarfs.

In our simulations, stars form in dense stellar clusters with high
gas-to-star conversion efficiencies and rather uniform
metallicities. A fraction of these clusters remain bound after the gas
is removed by feedback, but others are destroyed, and their stars,
which typically have velocity dispersions of 20 to 40~km~s$^{-1}$,
expand until they become bound by the dark matter halo. We thus
speculate that the stars in ultra-faint dwarf galaxies may show
kinematic and chemical signatures consistent with their origin in a
few distinct stellar clusters.  On the other hand, some globular
clusters may form at the center of primordial dwarf galaxies and may
contain dark matter, perhaps detectable in the outer parts. 
\end{abstract}

\section{Introduction} \label{sec:intro}

The epoch of formation of the first stars and galaxies is poorly known
due to lack of direct observations and the difficulty of ab-initio
theoretical modeling. However, understanding this short cosmic epoch
is of great importance, not only on its own merit, but also to make
progress in other fields of research in astrophysics such as
near-field cosmology \citep[\eg,][]{RicottiGnedin2005,
  BovillRicotti2011b, Wheeler2015}, the epoch of reionization
\citep[\eg,][]{Gnedin2014,Wise2014,OShea2015} and the formation of the
progenitors of supermassive black holes \citep[\eg,][]{Volonteri2010,
  Katz2015}.

It has been established that the first stars in the universe
(Population~III) have unique properties because they formed in gas of
primordial composition, thus devoid of important coolants such as
carbon and oxygen, and with inefficient H$_2$ formation because of the
absence of dust. Simulations of the formation of Population~III stars
\citep{BrommCL1999, Abel2000, Turk2009} have shown that gas condenses
at the center of minihalos of mass $10^5$-$10^6$~M$_\odot$, reaching
densities of the order of $10^{11}$~cm$^{-3}$ on 100~AU
scales. Initially the density of the gas increases, driven by the
gravitational potential of the dark matter. As the gas becomes fully
molecular, the gas cools further and becomes self-gravitating forming
a proto-star \citep{Abel2002}. In order to numerically resolve the
starting phases of star formation in metal-free and dust-free
molecular clouds it is therefore necessary to achieve a numerical
resolution of a few 100~AU (or 1~M$_\odot$ mass scale).  As the first
stars synthesize and eject metals into the intergalactic medium (IGM)
and interstellar medium (ISM), the metallicity and dust-to-gas ratio
in star forming regions increase and the enhanced cooling rates allow
molecular clouds to form at lower mean density, and on larger mass and
spatial scales. Thus, the numerical requirements for resolving star
formation in molecular clouds become less stringent with increasing
gas metallicity \citep[\eg,][]{Bromm2001, Kuhlen2013, Tomassetti2015}.

Two approaches are widely used for modeling the formation of the first
dwarf galaxies in cosmological simulations. In the first approach
metal enrichment is calculated self-consistently resolving the
formation of the first stars at $z>10$ in relatively small (1-4
comoving Mpc$^3$) cosmological volumes \citep{Ricotti2002a,
  Ricotti2002b, WiseAbel2007, Ricotti2008, Muratov2013a, Muratov2013b,
  Wise2014}. The second approach, typically used in ``zoom
simulations'' of dwarf galaxies, adopts a metallicity floor (typically
$Z \sim 10^{-3}$~Z$_\odot$) everywhere in the IGM in order to initiate
normal Population~II star formation avoiding to capture Population~III
star formation \citep{Gnedin2009, Tassis2012, Christensen2012,
  Kuhlen2013, Hopkins2014, Thompson2014, Wheeler2015}

It is not difficult to suppress Population~II star formation in dwarf
halos. First, photo-heating during reionization shuts off accretion
below a characteristic mass $M_c(z)$, and internal ionization sources
can unbind the majority of baryons altogether \citep{Efstathiou1992b,
  Barkana1999, Gnedin2000, Hoeft2006, OkamotoGaoTheuns2008,
  Simpson2013, Benitezllambay2014, Ricotti2009, Pawlik2009,
  Sawala2014b}.  Second, slow formation of H$_2$ in pristine
almost-dust-free gas can also inhibit star formation.  Simulations
that model molecular chemistry and form stars in molecular clouds
confirm that dwarf halos can be left almost totally dark because gas
fails to collapse to sufficient density for dust to form, and cooling
to continue \citep{Gnedin2010, Kuhlen2012, Kuhlen2013, Jaacks2013,
  Thompson2014}. However, not surprisingly in light of the discussion
above, a number of authors have found that results for low-metallicity
dwarf galaxies depend on the chosen numerical resolution when star
formation is restricted to resolution elements in which the gas is
fully molecular \citep{Hopkins2012H2, Wise2014, Kimm2014,
  Tomassetti2015}. Thus, adopting such a sub-grid recipe in
simulations focused on the transition between Population~III to
Population~II stars may lead to an artificial suppression of star
formation in gas with metallicity below a critical threshold set by
the numerical resolution.

In this paper, we present new radiation-hydrodynamic cosmological
simulations of the formation of the first stars and dwarf galaxies in
which we adopt a model for Population~III and Population~II star
formation (\ie, without a metallicity floor) and their radiative and
mechanical feedback in small volume simulations (about 1 comoving
Mpc$^3$). We use the adaptive refinement tree (ART) code with
appropriate modifications as described in \S~\ref{sec:sim} and in more
detail in a companion paper (Parry, Ricotti \& Gnedin, in
preparation). While we cannot resolve all stages of star formation, we
do resolve dense clumps of gas on $0.1-1$~pc scales that (if Jeans
unstable) inevitably collapse into Population~III or Population~II
stars.  Contrary to other implementations of Population~II star
formation, we do not require the gas to be fully molecular but we set a
very high density threshold for star formation by requiring the gas to
be self-gravitating and converging at the maximum refinement level
(see \S~\ref{sec:sim}). With this choice we avoid the
metallicity-dependent resolution requirements for the gas to become
fully molecular and treat consistently the transition from metal-free
to metal-poor star formation. Star formation only takes place on the
maximum refinement level on parsec or sub-parsec scales with
efficiency $\epsilon_*$ of conversion of $\rho_{gas}$ into $\rho_*$ on
a local dynamical timescale (we explore a range for $\epsilon_*$
between 1\% and 100\%). We find that at sub-parsec resolution the
majority of Population~II stars (defined to have metallicity $Z >
10^{-5}$~Z$_\odot$) form in gas that is $10\%$ to $60\%$ molecular,
while metal-free gas forming Population~III stars is only partially
molecular at these scales.

The transition from Population~III to Population~II star formation has
been the focus of previous semi-analytical
\citep[\eg,][]{Scannapieco2003, Yoshida2004, Schneider2006} and
numerical simulations \citep[\eg,][]{Tornatore2007, Maio2010,
  Greif2011, Johnson2013, Wise2012a, Wise2012b, Muratov2013a,
  Muratov2013b, Wise2014}. This transition is important for predicting
whether the James Webb Space Telescope (JWST) will be able to observe
galaxies dominated by Population~III stars \citep{Pawlik2011,
  Zackrisson2011}, understanding the sources of IGM reionization
\citep{Ciardi2000, Ricotti2002b, RicottiOstriker2004b, Gnedin2008a,
  RicottiOM2008, MBK2014}, determining the origin of ultra-faint dwarf
(UFD) galaxies in the Local Group and probing dark matter and gravity
on small scales \citep{PolisenskyRicotti2011}.

Galaxies with masses below a critical value $M_c(z)$ are expected to
have an early truncation of their star formation histories due to
reionization or internal feedback mechanisms, and thus can be used as
a laboratory to investigate the conditions and star formation in the
high redshift universe \cite{BovillRicotti2011b, MBK2015}. These
systems are commonly referred to as ``fossil galaxies''
\citep{RicottiGnedin2005, BovillRicotti2009}, reflecting the idea that
their stellar populations are $> 11-12$~Gyr old. The best candidates
for such systems are UFDs (L$ < 10^5$~L$_\odot$) discovered in the past
decade \citep{Belokurov2007a, Belokurov2010, Koposov2015, DES2015} and
some old globular clusters (GCs) \citep{Ricotti2002, KatzR2013,
  KatzR2014}. Most UFDs contain only old, metal-poor stars and appear
to have simple star formation histories \citep{Brown2012, Brown2014},
making them excellent candidates for probing chemical and dynamical
signatures of the first generations of stars.

The focus of this paper is on how morphology begins to develop in the
earliest galaxies, before the epoch of reionization, as well the
relationship between morphology and the kinematic properties of the
stars and gas, their metallicities and modes of star formation.  In
particular, we are interested in establishing observational links
between simulated fossils galaxies and dwarf spheroidals and UFDs in
the Local Group for which detailed properties (including morphology,
kinematics and chemistry) are available, or can be probed by targeted
observations.  In previous works \citep{RicottiGnedin2005,
  BovillRicotti2009}, we found that the simulated fossils have stellar
spheroids with half-light radii of about $100$~pc (nearly independent
of their luminosity and mass-to-light ratio) and surface brightnesses
consistent with the faintest dwarf spheroidals and UFDs. Here, we
re-examine this question using simulations that have much higher
spatial resolution (about a factor of ten higher) and that, by
allowing star formation on sub-parsec scales, can resolve the eventual
formation of compact star clusters. These simulations are similar to
the ones presented in \cite{Muratov2013a}, as they are run with ART
and adopt similar (but not identical) sub-grid recipes for star
formation and feedback.  Another difference in our simulation with
respect to previous works \citep[\eg,][]{Wise2012, Wise2014} is that
we use a star formation recipe and radiation transfer methods (OTVET)
that allows the formation of hundreds to several tens of thousands of
``star particles'' per galaxy, with masses as low as $\sim
40$~M$_\odot$. We also form stars stochastically, checking for
suppression of star formation due to feedback with time resolution of
$10^5$~years. The masses of the stellar particles are not small enough
for sampling a realistic IMF, but they are much smaller than a cluster
mass and allow us to resolve the formation of compact star clusters
that may remain bound if their star formation efficiency is close to
$50\%$ \citep{Hills1980, GeyerB2001}.  This is of crucial importance
because in the local universe star formation is observed to occur only
in clusters \citep{Lada2003} and likely the same is true at high
redshift \citep{Clark2008, Karlsson2012}. However previous
cosmological simulations, (especially the ones that use ray-tracing
for radiative transfer) could not afford to form individual stars, so
instead employed stellar particles representing whole clusters of
stars.

The numerical improvement described above allows us to revisit
important questions on the morphology an chemical signatures in the
first galaxies:
\begin{itemize}
\item ``Emergence of the Hubble sequence'': Is there a genuine trend for more irregular
and spheroidal galaxies at high-z or vice versa gas-rich disks are common?
\item ``Origin of dSphs vs dIrrs'': Are the morphologies of dSphs and UFDs in the local
  group set at formation or are the result of tidal interactions with the Milky-Way?
\item ``Size and surface brightness of fossil galaxies'': What
  determines the half-light radii and surface brightnesses of dSphs and
  UFDs?
\item ``Unified theoretical model for the formation of compact stellar clusters and
  UFDs'': The discovery of UFDs and dwarf-globular transition objects
  \citep[\eg,][]{Willman2012, Forbes2013} somewhat blurred the
  distinction between compact stellar clusters and dwarf galaxies. Is
  there a deeper link between these objects?
\end{itemize}
The layout of this paper is as follows.  In \sect{sec:sim} we briefly
describe the cosmological code ART used in this work, initial
conditions and the physics included in the simulation, as well as the
modifications we have made to model metal-free star formation.  Our
results are presented in \sect{sec:res} and a discussion in
\sect{sec:disc}. Summary and conclusions are in \sect{sec:conc}.
\begin{deluxetable*}{lcccccccccccc}[ht]
\tabletypesize{\scriptsize}
\tablecaption{The numerical parameters adopted for each of our simulations.\label{tab:ics}}
\tablewidth{0pt}
\tablehead{
\colhead{Label} &
\colhead{$\Delta x$} &
\colhead{M$_{\rm DM,part}$} &
\colhead{N$_{\rm levs}$} &
\colhead{M$_{\rm *, min}$} &
\colhead{Z$_{\rm crit}$} &
\colhead{n$_{\rm H,pIII}$} &
\colhead{f$_{\rm H_{2},pIII}$} &
\colhead{E$_{\rm SN,pIII}$} &
\colhead{E$_{\rm SN,pII}$} &
\colhead{IMF} &
\colhead{$t_{\rm *,samp}$} &
\colhead{$\epsilon_*$}\\
\colhead{} &
\colhead{com. pc} &
\colhead{[$10^{3}$~M$_{\odot}$]} &
\colhead{} &
\colhead{[M$_\odot]$} &
\colhead{[Z$_\odot$]} &
\colhead{[cm$^{-3}$]} &
\colhead{[$10^{-5}$]} &
\colhead{$[10^{51}$~ergs]} &
\colhead{$[10^{51}$~ergs]} &
\colhead{(Pop~II)} &
\colhead{[Myr]} &
\colhead{}
}
\startdata
REF\tablenotemark{a}   & 10.9 & 49.2 & 10 & 40 & $10^{-5}$ & 1.0 & 1.0 & 30 & 1.0 & Chab. & 0.1 & 0.1\\
HSFE\tablenotemark{a}  & 10.9 & 49.2 & 10 & 40 & $10^{-5}$ & 1.0 & 1.0 & 30 & 1.0 & Chab. & 0.1 & 1.0\\
LSFE\tablenotemark{a}  & 10.9 & 49.2 & 10 & 40 & $10^{-5}$ & 1.0 & 1.0 & 30 & 1.0 & Chab. & 0.1 & 0.01\\
AM13-REF\tablenotemark{b} & 10.9& 5.5 & 9 & 40 & $10^{-5}$ & 1.0 & 1.0 & 30 & 1.0 & Chab. & 0.1 & 0.1\\
\enddata
\tablenotetext{a}{IC with cosmological parameters:
$(\Omega_m, \Omega_\Lambda, \Omega_b, \sigma_8, n_s)=(0.30, 0.70, 0.040, 0.90, 1.00)$}
\tablenotetext{b}{IC with cosmological parameters:
$(\Omega_m, \Omega_\Lambda, \Omega_b, \sigma_8, n_s)=(0.28, 0.72, 0.046, 0.817, 0.96)$} 
\end{deluxetable*}
\section{The Simulations} \label{sec:sim}
 
Our simulations follow the evolution of an ensemble of galaxies in a
1\hinvmpc volume of the Universe prior to the epoch of reionization.
In practice, the end point of the simulations is $z \sim 9$.  The
simulation code \art\citep[Adaptive Refinement
  Tree;][]{Kravtsov1997,Kravtsov1999,Kravtsov2003,Rudd2008}, employs
an Eulerian scheme to track the dynamics of gas, dark matter and
stars, adaptively improving the spatial and temporal resolution in the
densest and most rapidly evolving regions of the simulation volume.
The propagation of ionizing photons emitted by young stars is tracked
through radiative transfer calculations that are self-consistently
coupled to the hydrodynamics.  The code also includes a prescription
for modeling the abundance of molecular Hydrogen (\hmol) that includes
H$^{-}$ catalyzed formation and formation on dust grains,
self-shielding and dust shielding
\citep{Gnedin2009,GnedinKravtsov2011}.  Full details of our version of
\art can be found in Section~2 of a companion paper (Parry, Ricotti,
Gnedin 2016, in preparation) (hereafter PRG16) and references therein.
The following subsections recap those parts of the code that are
particularly relevant to dynamics and morphology in the first
galaxies.

Three-dimensional radiative transfer in four energy bands
is solved and coupled to the hydrodynamics calculation using the the
OTVET approximation \citep{Gnedin2001}. We consider the
propagation of \HI, \HeI and \HeII ionizing photons, as well as \hmol
dissociating photons in the Lyman-Werner bands.  Self-shielding and
dust-shielding of \hmol are included using an observationally
motivated model.  Both shielding factors are computed as functions of
the local column densities of \HI and \hmol, which are approximated as
$n\rho/|\nabla\rho|$.  \art keeps track of a non-equilibrium chemical
network that includes five species of atomic and ionized Hydrogen and
Helium, as well as molecular Hydrogen.  The abundances of each
species, together with the local UV radiation intensity, are used to
compute self-consistent heating and cooling rates.

\subsection{Initial Conditions} \label{sec:ics}

The simulations presented here begin from two different sets of
initial conditions, both of which have formed part of previously
published works; \citet{Ricotti2002a,Ricotti2002b} (hereafter R02) and
\citet{Muratov2013a,Muratov2013b} (hereafter AM13).  Both assume a
\lcdm cosmology and represent a cubic volume of the Universe,
1\hinvmpc on a side.  The properties of the two sets of initial
conditions are listed in \tab{tab:ics}.

In all of our simulations the number of dark matter particles, $N_{\rm
  DM}$, is equal to the number of cells in the root level mesh $N_{\rm
  root}$, which sets the particle mass, $M_{\rm DM} = \rho_{\rm
  crit}\Omega_{DM}V_{\rm sim} /N_{\rm root}$, where $\rho_{\rm crit}$
is the critical density and $V_{\rm sim}$ is the simulation volume.

\subsection{Mesh Refinement}

Three criteria control the refinement of the simulation mesh.  Cells
are refined when their gas or dark matter masses exceed the threshold
values $M_{\rm gas,th}$ and $M_{\rm DM,th}$. We set $M_{\rm DM,th}$
equal to the mass of one dark matter particle and the gas mass
threshold is then fixed at $M_{\rm DM,th}\times \Omega_{b}/\Omega_{\rm
  DM}$, such that the two refinement criteria are equivalent for a
cell with the cosmic baryon fraction.  In addition, we require that
the Jeans length of the gas, $L_{\rm J}=20.7~{\rm pc} (T/n)^{1/2}$, be
resolved by at least 5 cell lengths, which satisfies the condition
described by \citet{Truelove1997} required to avoid artificial
fragmentation.

\subsection{\popII Star Formation and Feedback}

\popII star formation is allowed only when
\begin{enumerate}
\item The cell is maximally refined: $\Delta x = 10$~comoving pc (see \tab{tab:ics}).
\item The flow is convergent ($\nabla.\vec{v} < 0$).
\item The Jeans's length of the gas
  can no longer be resolved by five cells ($L_{\rm J}< 5~{\Delta
    x}$). For the value of $\Delta x$ in our
  simulations $L_{\rm J}< 50$~comoving pc, translates in the density
  threshold for star formation: 
\[
n_{\rm H}>n_{\rm H, pII} \approx 1.7 \times 10^3~{\rm cm}^{-3} \left(\frac{T}{100~K}\right)
  \left(\frac{1+z}{10}\right)^2. 
\]
Phase diagrams show that the bulk of Population~II star formation
takes place at densities $10^4-10^5$~cm$^{-3}$, with maximum gas
densities as high as $10^6$~cm$^{-3}$ in few cells. The star forming
cells have temperatures ranging $T \sim 10-10^3$~K and $f_{H_2} \sim
10\%-60\%$.
\item We also require that the gas overdensity is at least $\delta_{\rm gas}>2000$ and the gas temperature $T<10^4K$.
\end{enumerate}
If all four conditions are met, star formation starts after one
dynamical time with rate:
\begin{equation} \label{eqn:sf}
\frac{d\rho_*}{dt}=\epsilon_*\frac{\rho_{\rm gas}}{t_{\rm dyn}}
\end{equation}
where $\rho_{\rm gas}$ is the gas density, $t_{\rm dyn}=(3\pi/32
G\rho_{\rm gas})^{0.5}$ is the dynamical time and $\epsilon_*$ is the
fraction of gas converted into stars in $t_{\rm dyn}$.  Note that -
similarly to the recipe for Population~III star formation discussed
below - we do not require the gas to be fully molecular
($f_{H_2}=100\%$).

Each \popII star particle represents a stellar population with a
\citet{Chabrier2003} initial mass function (IMF).  They can lose mass
through stellar winds associated with massive stars and through the
ejection of material by supernovae (SNe).  Stellar particles emit
radiation with the spectral energy distribution (SED) shown in figure
4 of \citet{Ricotti2002a} and an overall normalization that evolves
with age according to the Starburst99 model of \citet{Leitherer1999}.
This results in a radiative output that begins to fall off rapidly
after 3~\Myr and has dropped by four orders of magnitude after $\sim
30$~\Myr. Stars are formed stochastically over a typical time scale of
$1$~\Myr and have a minimum mass of M$_{\rm
  *,min}=40$~M$_\odot$. Conditions for star formation are checked every
$t_{\rm *, samp}=0.1$~\Myr.  SNe explosions begin after a time delay
equal to the lifetime of a 8~M$_\odot$ star ($3.4$~Myr) and continue
for a total of $35$~Myr, with each SN generating $E_{\rm
  SN,pII}=10^{51}$~ergs of thermal energy and ejecting overall a mass
in metals equal to 1.1\% of the initial stellar mass.  At each time step
over this interval, the total energy and metals produced by the stellar
population are deposited in the host cell.

\subsection{\popIII Star Formation and Feedback} \label{sec:pop3}

To be eligible for \popIII star formation, cells with metallicity less
than $Z_{\rm crit}=10^{-5}$~Z$_\odot$ must have gas densities and molecular fractions
exceeding threshold values:
\begin{enumerate}
\item $n_{\rm H} > n_{\rm H,pIII}=1.0$~cm$^{-3}$,
\item \fhmol$> f_{\rm H_{2},pIII}=10^{-5}$.
\end{enumerate}	
When these conditions are satisfied, a particle representing a single
\popIII star with mass $M_{\rm pIII}=40$~M$_\odot$ is formed during
the next time step.  Each \popIII star has a lifetime of 3.9~\Myr and
radiates with the same SED used for \popII star particles, but with an
increased ionizing luminosity, in agreement with the model from
\cite{Schaerer2002} for a zero metallicity star of 40~M$_\odot$.  At
the end of their lives, \popIII particles explode as hypernovae
\citep{UmedaNomoto2003} and eject $8$~M$_\odot$ of metals and $E_{\rm
  SN,pIII}=30 \times10^{51}$~erg of thermal energy into their
immediate surroundings.  Following \citet{Muratov2013a,Muratov2013b},
we distribute metal and energy evenly over a sphere of radius $1.5$
cell lengths around the star.  The assumed thresholds for star
formation are also listed in \tab{tab:ics}.

\subsection{Analysis} \label{sec:analysis}

Unless otherwise specified the results shown in the next sections
refer to galaxies identified at redshift $z=9$.  We analyze the
simulations using two forms of data - log files that record global
properties of the simulation at every root-level time step, and output
snapshots containing the properties of the dark matter, stars, gas and
radiation field throughout the simulation volume.  Snapshots are
written at values of the expansion factor separated by 0.01 until
0.05, and by 0.005 thereafter.

Individual galaxies are identified using a modified version of the
\subfind halo finder code \citep{Springel2001}.  The first phase of
the algorithm identifies ``friends-of-friends" groups
\citep{PressDavis1982,Davis1985} by linking dark matter and star
particles separated by less than $0.2$ times the mean inter-particle
separation.  It then finds gravitationally bound substructures
(sub-halos) by iteratively unbinding particles around local density
peaks.  The center of each galaxy is deemed to be the potential
minimum within the sub-halo. Although \subfind identifies all
substructures with at least 32 particles, we impose a more
conservative limit of 50, consistent with \citet{Kravtsov2004} who
found that the cumulative mass function was converged above that
resolution threshold.  The minimum galaxy mass considered in the
following sections is therefore \solmass{2.5}{6} and \solmass{2.8}{5}
for the \RLREF and \AIREF simulations respectively. We stress that,
while the halo mass function can be considered substantially complete
down to this limit, the star formation histories of halos close to the
limit will likely be affected by a lack of resolution in their
progenitors.  Where merger trees are required for the analysis, they
are constructed by linking sub-halos in each snapshot with any
progenitor in the previous snapshot that contains at least five per
cent of their dark matter particles.
\begin{figure*}[t]
\begin{center}
\includegraphics[width=17.5cm]{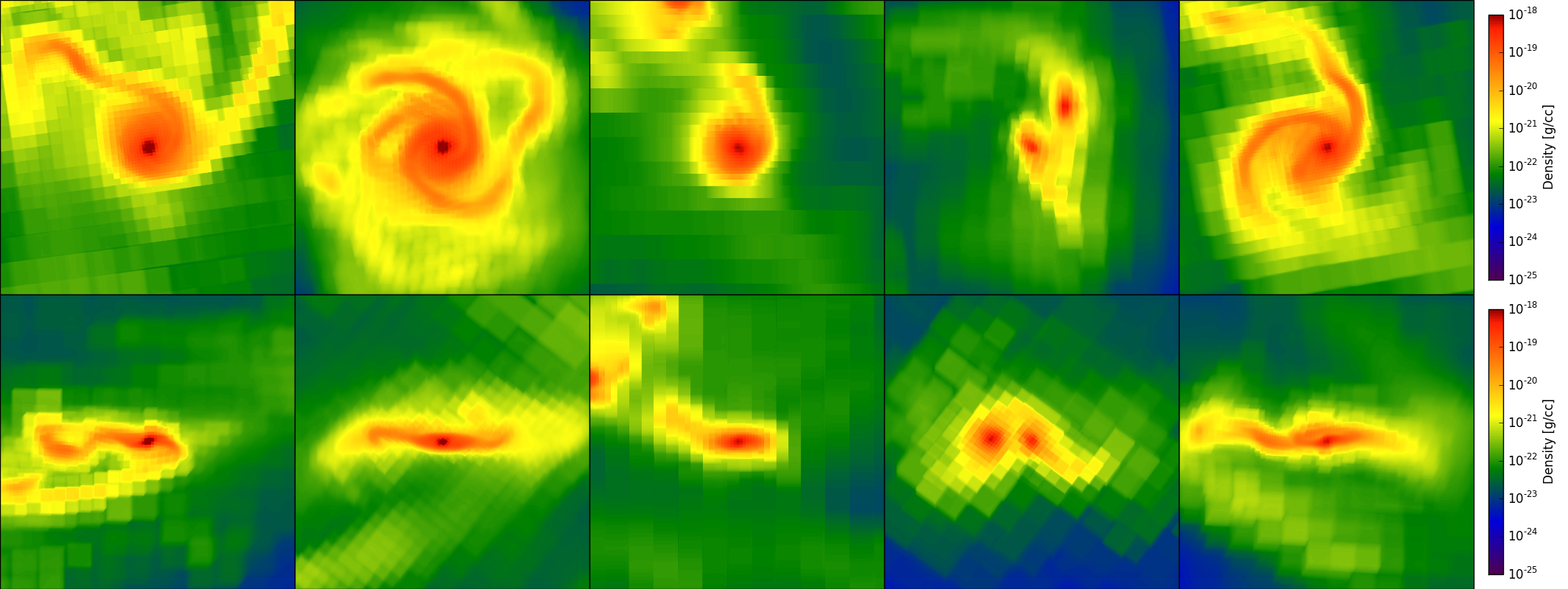}
\vspace{0.05cm}
\caption{Projected gas density in the top five galaxies ranked by
  stellar mass at $z=9$.  In the top row the projection direction is
  parallel to the net angular momentum vector of all gas within 2 per
  cent of the virial radius; in the bottom row, the projection
  direction is normal to that vector.  Each image is 100~pc on a
  side. Image produced using the yt python libraries, \citep{yt2011}.} 
\label{fig:gas_images}
\end{center}
\end{figure*}
\begin{figure*}[t]
\begin{center}
\includegraphics[width=17.5cm]{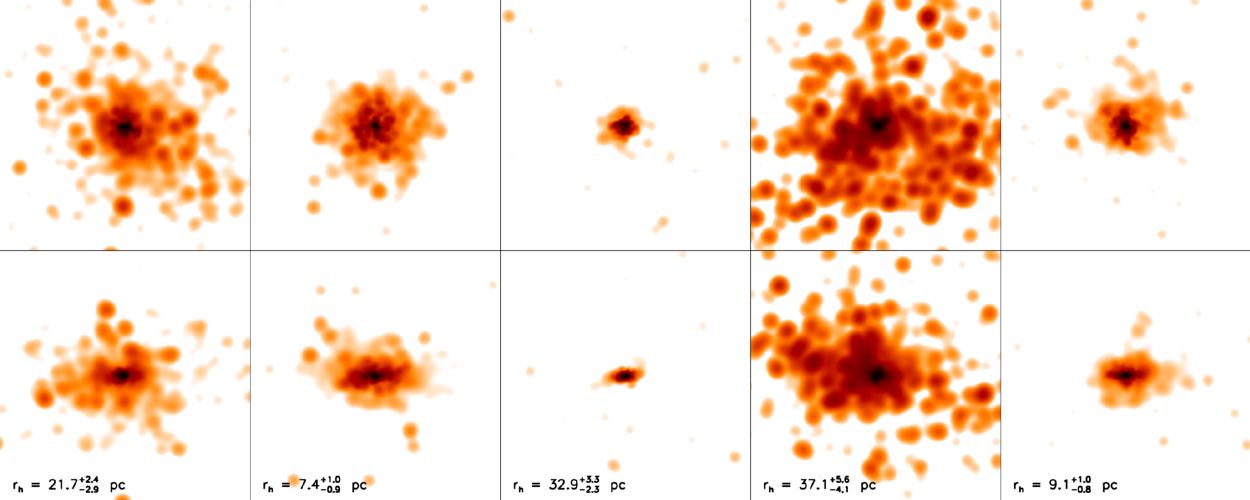}
\vspace{0.05cm}
\caption{Images of the smoothed projected stellar density in the top
  five galaxies ranked by stellar mass (counterparts to the images in
  \fig{fig:gas_images}) at $z=9$.  Details of the smoothing technique can be
  found in the text.  In the top row, the projection direction is
  parallel to the net angular momentum vector of all gas within two per
  cent of the virial radius; in the bottom row, the projection
  direction is normal to that vector.  Each image is $100$~pc on a side.
  Labels indicate the projected half-mass radius of the stars in each
  galaxy.}
\label{fig:star_images}
\end{center}
\end{figure*}

\section{Results}\label{sec:res}

\subsection{Gas Disks and Stellar Spheroids}\label{sec:morph}

\fig{fig:gas_images} shows that gas disks, although rather thick, are
clearly identifiable in many of the brightest (top five ranked by
stellar mass) galaxies at redshift $=9$.  The projected gas density is
shown in two orientations for each galaxy, parallel and normal to the
angular momentum vector of the gas.  In the cases where a well-defined
gas disk is present, this gives face-on and edge-on views.

In \fig{fig:star_images} we show the projected stellar mass density
for the same five galaxies that appear in \fig{fig:gas_images}.  The
mass from each star particle is spread out over nearby pixels using an
SPH-like kernel\footnote{Letting u = r/h, where r is the (2D)
  displacement from the kernel center and h is the smoothing length,
  the kernel has the form $w(u) = w_0 + w_1(u-1)u^2$ for $u<0.5$ and
  $w(u) = w_3(1-u^3)$ for $u\geq0.5$.  The constants $w_0$, $w_1$ and
  $w_2$ are chosen such that $\int_0^1 w(u) \mathrm{d}u = 1$.}
enclosing 32 neighbors.  The projection axes and scales of the images
are the same as in \fig{fig:gas_images}.  The morphology of the
stellar component of these galaxies is clearly much closer to a
spheroid than that of the gas, although some flattening is apparent in
the same sense.

\subsection{Circularity of Stellar Orbits}

A straightforward way to identify stellar disks in simulated galaxies
is to compute the circularity ($\mathcal{E}$) of star particle orbits
\citep[e.g.][]{Abadi2003,Aquila}.  In a coordinate system where the net
angular momentum of all of the galaxy's stars is in the positive z
direction, circularity may be defined as:
\begin{equation}
\label{eqn:CIRC_DEF}
\mathcal{E}_{\rm E} = \frac{J_z}{J_{\rm circ}(E)},
\end{equation} 
where $J_z$ is the $z$ component of the star's specific angular
momentum and $J_{circ}(E)$ is the specific angular momentum of a star
with the same binding energy on a circular orbit.  For an infinitely
thin, rotationally supported disk, the distribution of circularities
is a $\delta$ function at $\mathcal{E}_{\rm E}=1$, while a
non-rotating, dispersion dominated spheroid gives rise to a broad,
symmetric distribution peaking at $\mathcal{E}_{\rm E}=0$.
\begin{figure}[t]
\begin{center}
{\includegraphics[width=8cm]{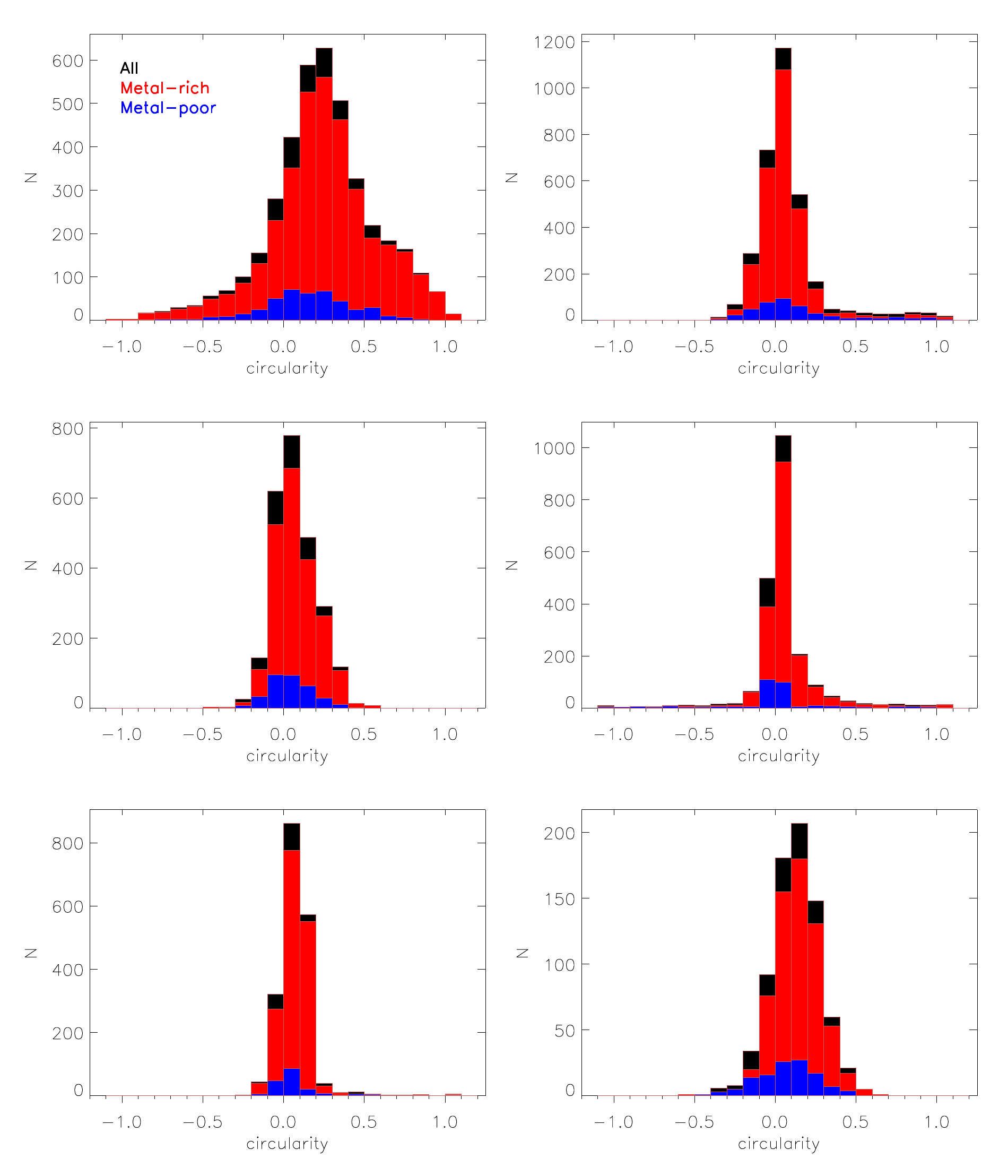}}
\vspace{0.05cm}
\caption{The distribution of orbital circularities (as defined in
  \eqn{eqn:CIRC_DEF}) for stars in the six brightest galaxies in our
  \RLREF simulation.  Blue and red histograms correspond to metal-poor
  and metal-rich stars respectively, divided at [Fe/H]=-1.5.}
\label{fig:circs}
\end{center}
\end{figure}

\fig{fig:circs} shows histograms of $\mathcal{E}_{\rm E}$ for all star
particles in the six brightest galaxies in our \RLREF simulation.
Most of the six distributions are consistent with non-rotating
spheroids, but in two cases the mean of the distribution is positive,
suggesting that the spheroid is rotating, or that a thickened disk
structure is superimposed on the non-rotating spheroid.  No obvious
difference is apparent between metal-rich stars ([Fe/H]$>-1.5$, red) and
metal-poor stars ([Fe/H]$<-1.5$, blue).  

\subsection{Size and Surface Brightness of Stellar Spheroids}
 	
\begin{figure}[t]
\begin{center}
{\includegraphics[width=8cm]{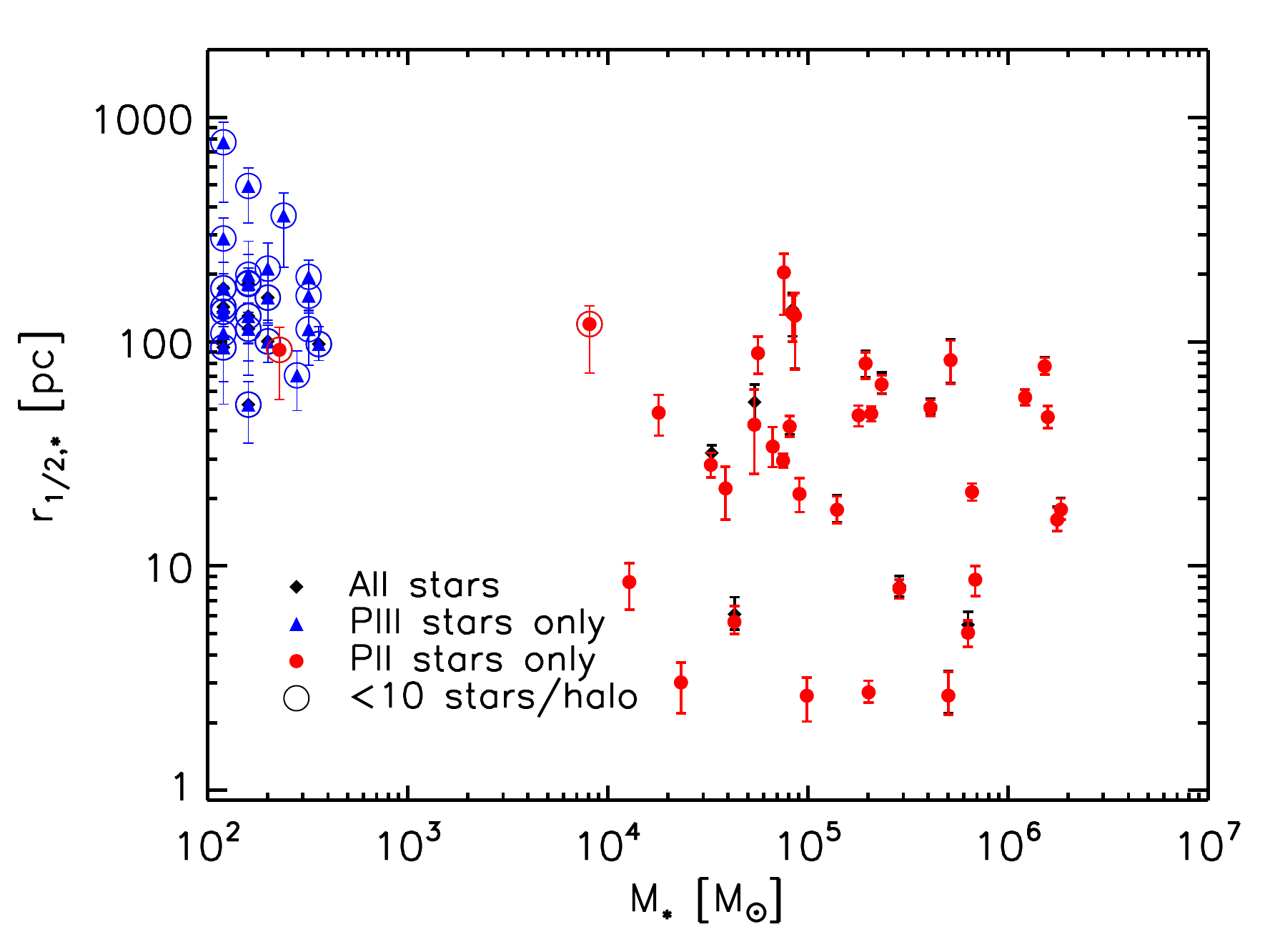}}
\caption{Stellar half-mass radius as a function of stellar mass for
  all galaxies with more than five star particles in the \RLREF
  simulation at $z=9$. The projected radii were computed in 100
  different random orientations; error bars indicate the 10$^{\rm th}$
  and 90$^{\rm th}$ percentiles of the distribution in each case.
The halos clustered in the upper left corner of the figure are halos
containing only Population~III stars. Halos typically obtain multiple
Population~III stars from accretion of multiple sub-halos containing
single Population~III stars with $M_*=40$~M$_\odot$.}
\label{fig:rh}
\end{center}
\end{figure}
\fig{fig:rh} shows the projected stellar half-mass radii (\rhalf) for
all galaxies in our \RLREF simulation with five or more star
particles.  To quantify the effect of different viewing angles, \rhalf
was computed 100 times for each galaxy using random orientations.
Error bars indicate the 10$^{\rm th}$ and 90$^{\rm th}$ percentiles of
the resulting distribution for each galaxy.  In contrast to Local
Group dwarfs, which show a relatively clear correlation between \rhalf
and stellar mass \citep[e.g.][]{Martin2008}, the simulated galaxies
can vary by a factor of 100 in \rhalf at a fixed stellar mass.  The
radius \rhalf is almost independent of viewing angle in many of the
galaxies, suggesting relatively spherical distributions, although, as
expected, the variation increases at lower $M_{*}$ where there are
fewer star particles per galaxy. Note that the points clustered at the
top left corner of the figure are not a numerical artifact but halos
that only contain \popIII stars (colored in blue). These halos
have an extended and low surface brightness stellar spheroid produced
by mergers of several minihalos containing \popIII
stars. However, if \popIII stars are indeed massive as we have
assumed here, the present-day fossil remnants of these objects would
be totally dark.

In order to make sure that the spatial distribution and kinematics of
the stars are not affected by numerical issues, we performed a series
of tests.  First, we estimated the stellar radii of the spheroids by
defining a radius containing half of the star particles, rather than
half of the stellar mass. Given that star particles in our simulations
have a wide range of masses, we
checked whether a few massive star particles that sank to the center
of the halo could be biasing the half-mass radii to smaller
values. Indeed, it makes sense that star particles, which represent a
collection of stars, should be treated as an extended distribution,
rather than a point mass. We found no significant differences between
the half-mass radius and the radius containing half of the star
particles in each case. Thus, the compact star clusters with half-mass
radii of a few pc in our simulations are not a numerical artifact, but
truly a compact collection of many star particles in virial
equilibrium (see below).
\begin{figure*}[th!]
\begin{center}
\includegraphics[width=8.5cm]{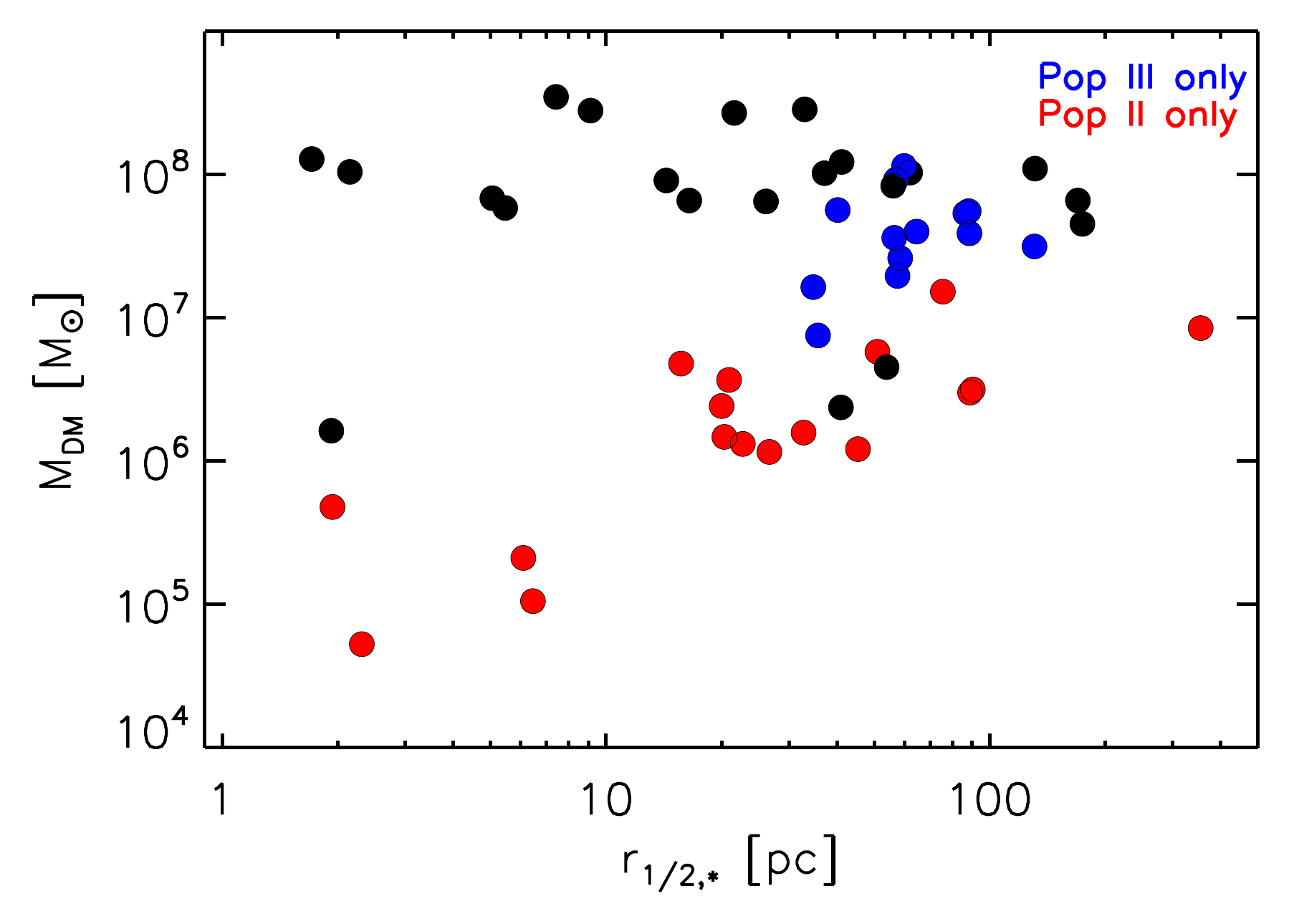}
\includegraphics[width=8.5cm]{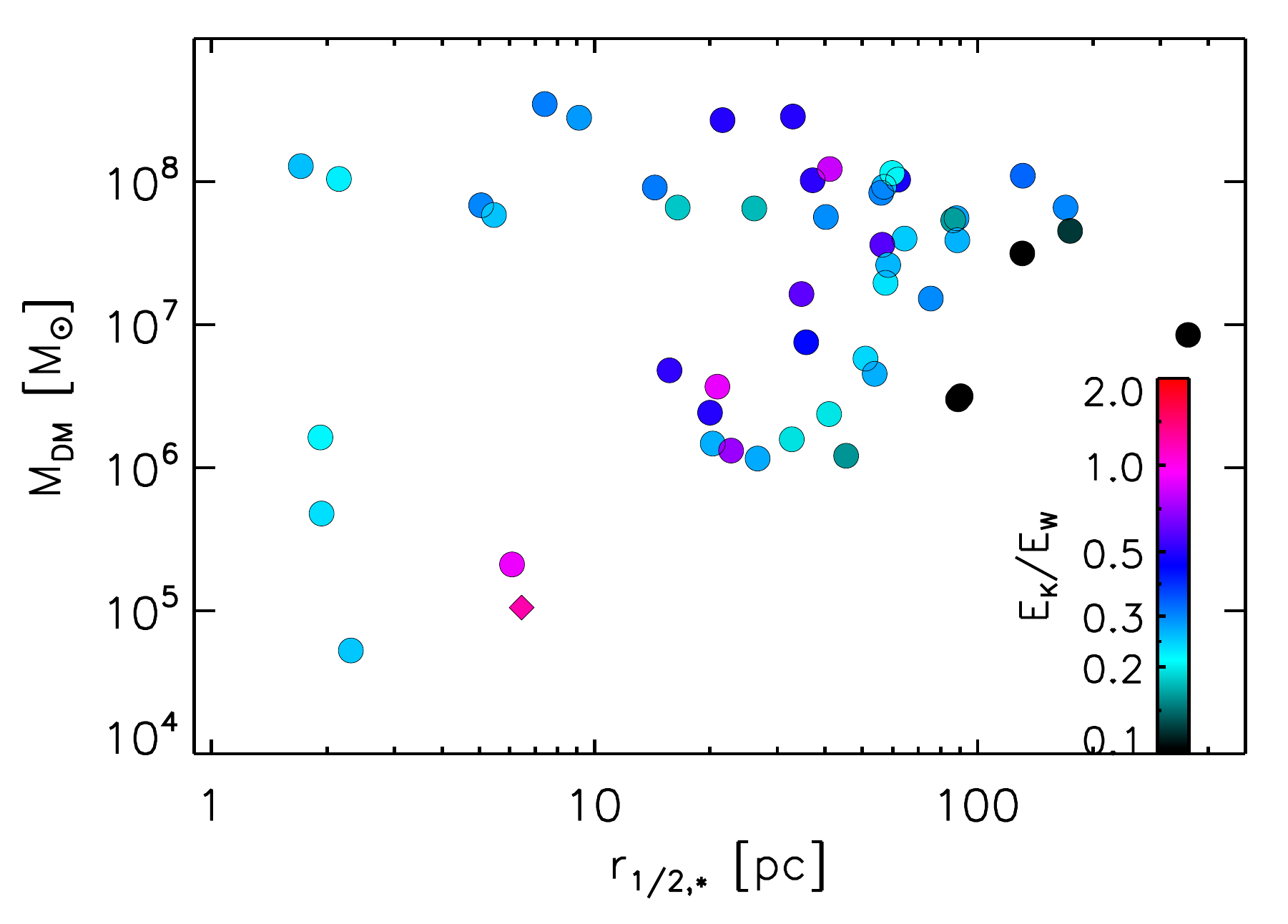}
\caption{({\it Left.}) Dark matter mass associated with galaxies or
  compact stellar clusters as a function of their stellar half-mass
  radius for all bound objects with more than five star particles in
  the \RLREF simulation. Dark matter halos containing both
  Population~II and Population~III stars are shown as black points,
  halos with only Population~III stars are shown in blue, and bound
  objects with only Population~II stars (pre-enriched by external SNe)
  are shown as red points. The small masses in dark matter
  ($10^5-10^7$~M$_\odot$) of the bound objects with only Population~II
  stars suggest the existence of a mode of triggered star formation in
  satellite halos of the first galaxies, leading to the formation of
  compact star clusters, possibly bound. ({\it Right}.) Same as
  the left panel but with color coding showing the ratio of the
  kinetic to gravitational binding energy of the system.}
\label{fig:Mdm_r}
\end{center}
\end{figure*}

The second test regards the kinematics of the stars. We wanted to
check whether the half-mass radii are different for lighter star
particles than for heavier ones. This would be the case if unphysical
2-body interactions between light star particles and dark matter or
massive stellar particles were to dynamically heat the system.  We
found that, excluding progressively more massive subsets of particles,
starting with the lightest, there was no obvious effect on the
calculated value of \rhalf.

\subsection{Properties of Compact Stellar Clusters and Low Surface Brightness Dwarfs}\label{sec:prop}

\fig{fig:Mdm_r}(left panel) shows the dark matter mass as a function
of the stellar half-mass radius for all galaxies with more than five
star particles in the \RLREF simulation. The color coding shows bound
objects that have only \popII stars (red), only \popIII (blue) and
both (black). In this plot we note a few interesting properties:
\begin{itemize}
\item {\it Normal dwarf galaxies:} These are bound objects with
  $M_{dm} \geq$\appmass{8} and contain both \popII and \popIII stars.
\item {\it Failed dwarf galaxies or ``dark galaxies'':} These are
  bound objects with $M_{dm} \sim$\appmass{7} that failed to form
  \popII stars (about 13 objects). They contain only \popIII stars
  distributed in a relatively extended stellar halo ($>50$~\pc). Their
  fossils today would be dark, unless low mass \popIII exist.
\item {\it Triggered star formation and compact stellar clusters:}
  These are bound objects with $M_{dm} \leq$ \appmass{6} that have
  \popII stars but do not have any \popIII star (about 13
  objects). Hence, either they are polluted with metals by galactic winds from nearby
  galaxies, or the \popIII star was ejected. These bound objects are
  likely an example of ``triggered'' star formation induced by star
  formation in more massive (\appmass{8}) halos \citep{Smith2015}.
Some of these objects are sufficiently compact to be
  candidate GCs. Their metallicity is also consistent
  with their identification as progenitors of today's old GCs.
\end{itemize}
Five of the 10 most compact objects ($r_h<7$~pc) are dark matter
dominated and in halos with $M_{dm}\sim$\appmass{8}. The other five
are baryon dominated with $M_{dm}<$\appmass{6}. Excluding halos
without \popII stars (that would be dark today), compact luminous
objects are found to have a bimodal mass distribution.

The right panel of \fig{fig:Mdm_r} is the same as the left except the
color coding shows the ratio of the kinetic to gravitational binding
energy of the stars in each object.  Galaxies with \rhalf$\sim
1-2$~\pc and \rhalf$\sim 50-200$~\pc appear to be bound ($|E_K/E_W| <
1$) and close to virial equilibrium ($|E_K/E_W| \sim 1/2$).  Objects
with intermediate radii, \rhalf$\sim 10-40$\pc, tend to be more
loosely bound and further from virial equilibrium ($|E_K/E_W| \sim
1-2$), suggesting that these stellar systems have not yet reached an
equilibrium configuration and that their half-mass radii might still
be expanding as a result of gas mass loss in young star-forming
clusters.
\begin{figure}
\begin{center}
\includegraphics[width=8cm]{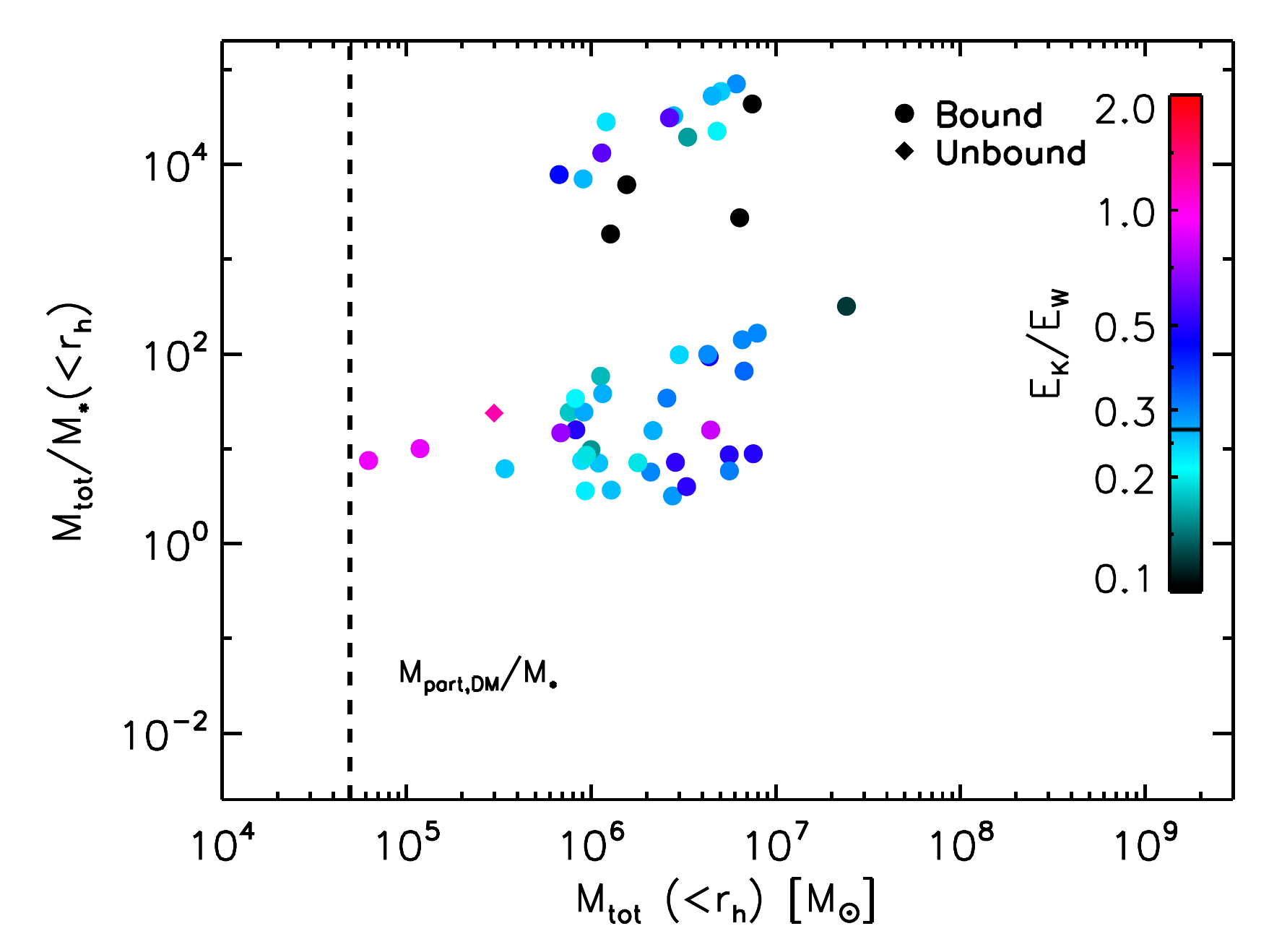}
\vspace{0.05cm}
\caption{Pseudo-mass-to-light ratio ($M_{\rm
  tot}(<\rhalfeqn)/M_{*}(<\rhalfeqn$) as a function of dynamical
mass $M_{\rm tot}(<\rhalfeqn)$ for all galaxies with more than five
star particles in the same simulation as in Fig.~\ref{fig:Mdm_r}.}
\label{fig:M-L}
\end{center}
\end{figure} 
\begin{figure*}[t]
{\includegraphics[width=17.5cm]{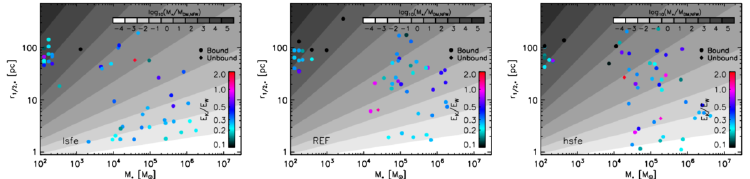}}
\caption{Galaxy sizes for three different values of the \popII star
  formation efficiency, $\epsilon_{*}$. See the text
  for an explanation of the colors and gray-scale regions.}
\label{fig:rh_sfe}
\end{figure*}

\fig{fig:M-L} shows the mass-to-light ratio $M_{\rm
  tot}(<\rhalfeqn)/M_{*}(<\rhalfeqn)$, (where $M_{\rm
  tot}(<\rhalfeqn)$ is the dynamical mass within \rhalf) as a function
of the total dynamical mass. The plot shows two separated set of
points: those with $M_{\rm tot}(<\rhalfeqn)/M_{*}(<\rhalfeqn)\sim
10^4$ are dark matter halos containing only \popIII stars (typically a
few stars as indicated by their total stellar masses of $M_* \sim
40-200$~M$_\odot$).  The second group of objects have either a
constant mass-to-light ratio of a few (the lower horizontal branch in
the plot) or a mass-to-light ratio that increases with dynamical mass
from ten to a few hundred (the upper branch in the plot). This
indicates that for a fixed dynamical mass of around $10^7$~M$_\odot$,
we either find dark matter dominated and low surface brightnesses
objects (analogous to the ultra-faint dwarfs), or compact star
clusters in which the dark matter is a subdominant component of the
dynamical mass, but which still can be embedded into larger dark
matter halos.

\subsection{Stellar Radii and Sub-grid Star Formation Efficiency}\label{sec:subgrid}

\fig{fig:rh_sfe} is similar to \fig{fig:rh} but for three simulations
adopting different values of $\epsilon_*$ (sub-grid star formation
efficiency) and with color coding of the points showing the ratio of
the kinetic to the gravitational binding energy of the objects
$|E_K/E_W|$. Each panel shows the stellar half-mass radii as a
function of stellar mass for all bound objects with more than five
star particles in the LSFE ($\epsilon_*=1\%$, left panel), \RLREF
($\epsilon_*=10\%$, center panel) and HSFE ($\epsilon_*=100\%$, right
panel) simulations.  The grayscale shaded regions show the stellar to
dark matter mass ratio within $r_h$, assuming for estimating the dark
matter mass within $r_h$ a NFW profile for a halo of mass \appmass{8}
and virial radius $2$~\kpc, that is rapresentative of typical halos at $z=9$
in $(1~Mpc)^3$ volume.  This is to illustrate that diffuse objects are
dark matter dominated while compact objects are baryon
dominated. Objects for which the ratio $|E_K/E_W|$ is greater than
unity (unbound systems) are represented by filled diamond symbols, all
other objects by filled circle symbols. We find qualitatively similar
bound objects in the three cases, independent of the assumed star
formation efficiency (even though it changes by a factor of 100
between the simulations). The spread of the \rhalf distribution for a
given mass (\ie, compact and low surface brightness galaxies) is found
in all the simulations, indicating that feedback effects determine the
interruption of star formation in the proto-star clusters. We do
observe, however, a weak dependence of galaxy luminosity on
$\epsilon_*$. The luminosity of dwarfs increases by about a factor of
two when $\epsilon_*$ is increased by a factor of 100. In the HSFE
simulation about 9-11 objects are as compact as GCs, and about 20 are
similar to UFDs, with low surface brightness and \rhalf$\sim
100-200$~\pc.

\subsection{Star Formation Histories and Metallicities}

In this subsection we investigate whether the compact dwarfs
resembling proto-GCs (based on morphology and kinematics), are also
consistent in terms of the metallicity distribution of their
stars. GCs are usually seen to have a rather uniform [Fe/H] abundance
in the range [Fe/H]=-2 to -1, indicative of a single stellar
population with little or no self-enrichment
\citep[\eg,][]{Grattonetal2004, Carrettaetal2009a}. Thus, to first
order approximation, their metallicity reflects the pre-enrichment of
their host galaxy.
\begin{figure}[th!]
{\includegraphics[width=8.5cm]{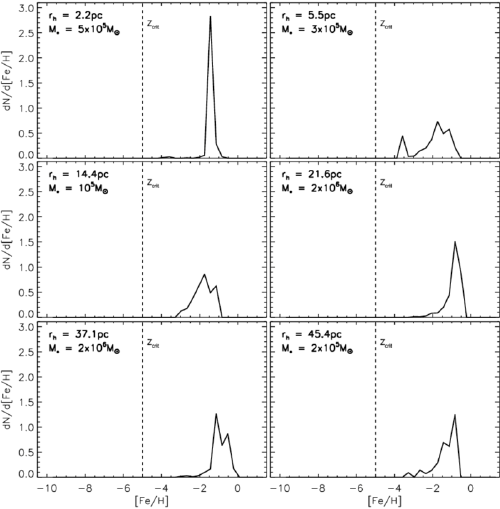}}
\caption{Distribution of star particle metallicities for six galaxies
  in the \RLREF simulation.  The panels are in increasing order of
  stellar half-mass radius, the value of which is included as a label,
  along with total stellar mass.}
\label{fig:met_dist_halos}
\end{figure}
In \fig{fig:met_dist_halos} we show the distribution of star particle
metallicities for six objects in the \RLREF simulation. Note that we
advect only one metallicity field, thus [Fe/H] is a proxi for the
total metal enrichment from SNe. The panels are ordered based on
stellar half-mass radii (from the smallest to the largest). The
stellar half-mass radius, \rhalf, is included as a label, along with
the total stellar mass. The dashed vertical lines indicate the
critical metallicity for transition from \popIII to \popII star
formation adopted in our simulations.  The most compact object in the
figure resembles today's old GCs: it has \rhalf$\approx 2.7$~\pc, $M_*
\approx $~\solmass{5}{5} and a delta-function-like metallicity
distribution peaked at [Fe/H]$=-1.5$.
\begin{figure*}[t!]
\begin{center}
	\includegraphics[width=17.5cm]{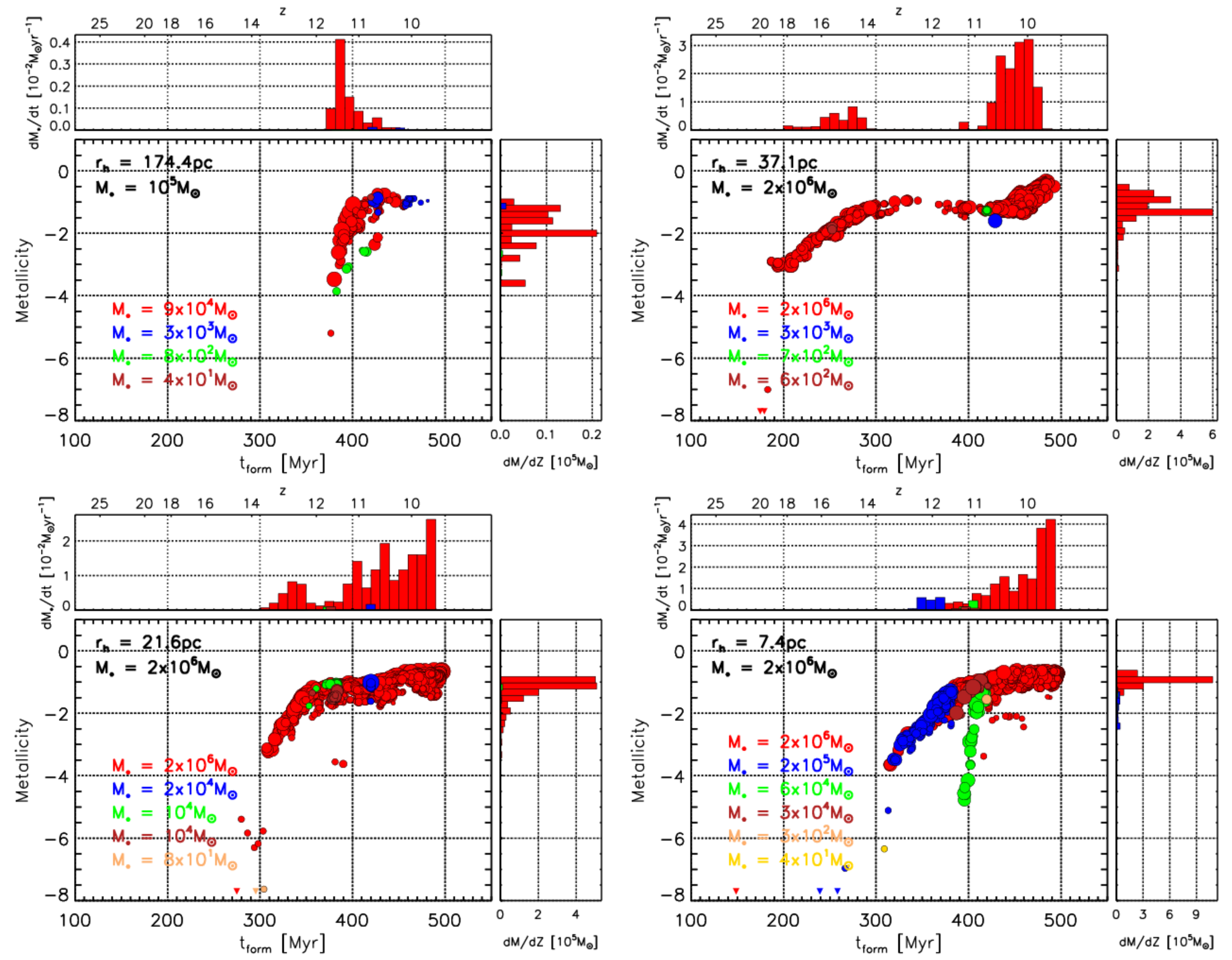}
\caption{The metallicities and formation times of all star particles
  belonging to four representative halos in the \RLREF simulation.
  Triangular points correspond to stars with metallicities below the
  lower limit of the ordinate axis.  Colors indicate the progenitor
  galaxy in which the star formed, red being the most massive
  (``main-branch") progenitor.  The labels in each panel indicate the
  stellar half-mass radius and stellar mass (top left legend) and the total
  mass formed in each progenitor (bottom left legend).  Histograms show the
  star formation rate and metallicity distribution. }
\label{fig:agemet_halos}
\end{center}
\end{figure*}
However, some compact stellar clusters in the simulation (for
instance, the one shown in the top-right panel of
\fig{fig:met_dist_halos}) have half-mass radii and stellar masses
consistent with old GCs, but much broader metallicity distributions,
with a range [Fe/H]$=-4$ to [Fe/H]$=-1$ and therefore it would be
classified as an ultra-compact dwarf. These are likely formed as the
result of the merger of at least three or four smaller star clusters
with different metallicities.  Thus, it is rather difficult to
distinguish between a GC and the nucleus of an ultra-compact dwarf galaxy
solely based on morphology. Indeed, some candidate proto-GCs in our
simulations are similar to the nuclei of compact dwarfs because they
form at the center of low mass dark matter halos.  Most simulated
dwarf galaxies with sizes $>10$\pc have rather broad metallicity
distributions, very similar to those observed in UFDs and classical
dSphs. We also find simulated dwarf galaxies with intermediate sizes
($15-50$\pc) and low stellar masses, much like the faintest UFDs found
around the Milky Way at distances of $\simlt 150$~\kpc.  Tidal
stripping of stars has been suggested as the reason for the small
sizes of these observed faint dwarfs \citep{BovillRicotti2011b}, but
our simulated dwarfs are already compact when they form, suggesting
that need not be the case.

Dwarfs with low stellar masses have metallicity distributions
characterized by several minor peaks, with the most prominent
typically at the highest metallicity. This suggests that the stellar
population is produced by in-situ formation or merger of a handful of small star
clusters with relatively narrow metallicity distributions. This
scenario is consistent with the rather bursty and spatially segregated
mode of star formation observed in most simulated galaxies. A smooth
and continuous mode of star formation and self-enrichment is rarely
observed in our simulations; a test of our model would be to compare it to
observations of the chemical and kinematic signatures typical of star
formation taking place in distinct star clusters
\citep{Bland-Hawthorn2015, Webster2016}.  Finally, even though the
critical metallicity assumed for the transition to \popII star
formation is [Fe/H]$=-5$, the lowest metallicities of \popII stars are
between [Fe/H]$=-4$ and [Fe/H]$=-3$. We have run some simulations with
different values of the critical metallicity (see PRG16) and have
found very little change in the distribution of stellar metallicities.

\fig{fig:agemet_halos} shows the metallicities and formation times of
stars in four galaxies in the \RLREF simulation.  While the star
formation histories of each galaxy are a complex function of many
factors, including merger history, halo mass and the local radiation
field, these four halos are broadly representative of the galaxy
ensemble as a whole.  Each color corresponds to an independent
progenitor galaxy which hosted the star particle when it formed.
Progenitor colors are ordered by the stellar mass contributed, as
indicated by the labels in the bottom left of each panel.  Star
particles were mapped to progenitor galaxies by determining their host
halo in the first output after they formed and then associating each
halo with a branch of the merger tree.  Merger trees were constructed
as described in \sect{sec:analysis}.  The stellar half-mass radius of
each galaxy is listed in the top left corner of each panel, along with
the total stellar mass.  The four examples shown suggest an inverse
correlation between stellar size and the fraction of stars formed late
in the simulation, which is borne out in the galaxy population as a
whole.  Each galaxy typically contains between two and eight \popIII
stars, with many progenitors forming just a single \popIII star.
Multiple "tracks" are evident in many of the most massive galaxies,
most obviously in the bottom right panel of \fig{fig:agemet_halos}.
The varying gradients of the tracks demonstrate different rates of
enrichment in independent progenitor halos.  We note that modifying
the parameter used to distinguish between \popIII and \popII stars
(\zcrit) affects the size of the ``gap" in metallicity between the
most metal rich \popIII star and the most metal poor \popII star, as
well as the time delay between the formation of the first \popIII star
and \popII star formation.  This delay occurs when gas is enriched
above \zcrit before it has reached a high enough density to form
\popII stars. For most choices of the simulation's free parameters the
delay has very little impact on the typical star formation and final
stellar mass of the galaxy, however a too large \zcrit or too strong
\popIII feedback can suppress \popII star formation completely in most
galaxies (see PRG16).

\section{Discussion: from Compact Stellar Clusters to Ultra-faint Galaxies} \label{sec:disc}

In this section we investigate the formation of stellar spheroids and
and what determines their sizes.  We have identified three processes that may play an important role: \renewcommand{\theenumi}{\roman{enumi}}%
\begin{enumerate}
\item The extent of the stellar distribution may simply reflect the
  extent of the star forming gas distribution. However, we found that
  the gas settles in disks with radii $50-150$\pc and a thickness of
  $10-20$\pc.  Thus, although all stars form in compact clusters
  within the disk, if the cluster becomes unbound it expands remaining
  confined by the gravitational potential of the dark matter halo
  (that to first approximation is spherical), thus a low surface
  brightness stellar spheroid with radius significantly larger than
  the disk thickness is formed (see below for a more quantitative
  discussion).
\item The stellar spheroid may increase in size due to repeated
  mergers that dynamically heat the system. This effect may play a
  role in increasing the size of an already extended stellar spheroid
  or stellar halo over a rather long time-scale (from $z=9$ to $z=0$)
  rather than self-gravitating compact clusters. Mergers appear to
  play a role in setting the size of the stellar spheroid in objects
  containing only \popIII stars (see objects at the top-left corned in
  Fig.~\ref{fig:rh}). Halos containing multiple \popIII stars have
  accreted them from mergers of minihalos containing single stars (we
  have check for this looking at different time snapshots tracing the
  merging of satellites). Because \popIII stars are accreted, their
  velocity dispersion is roughly the virial velocity of the host dark
  matter halo, or slightly lower due to the effect of dynamical
  friction. Thus their radial distribution extends to nearly $r_{\rm
    max}$.
\item In our simulations stars form in dense clusters with 3D velocity
  dispersion at formation in the range $20-40$\kms.  If the gas clump
  in which the cluster forms converts into stars $<50$\% of the gas,
  the cluster becomes unbound after the unused gas is expelled by SN
  feedback \citep[see,][]{Hills1980}. Expanding stars with a final
  velocity dispersion less than $v_{\rm cir} \sim 20$~\kms remain
  bound inside the dark matter halos and can extend close to $r_{\rm
    max}$. \label{item:clusterSF}
\end{enumerate}
The large range of radii observed in our simulations and their
weak dependence on the SFE, $\epsilon_*$, suggest that
\ref{item:clusterSF}) plays a dominant role.
\begin{figure*}[t!]
\begin{center}
\includegraphics[width=8.5cm]{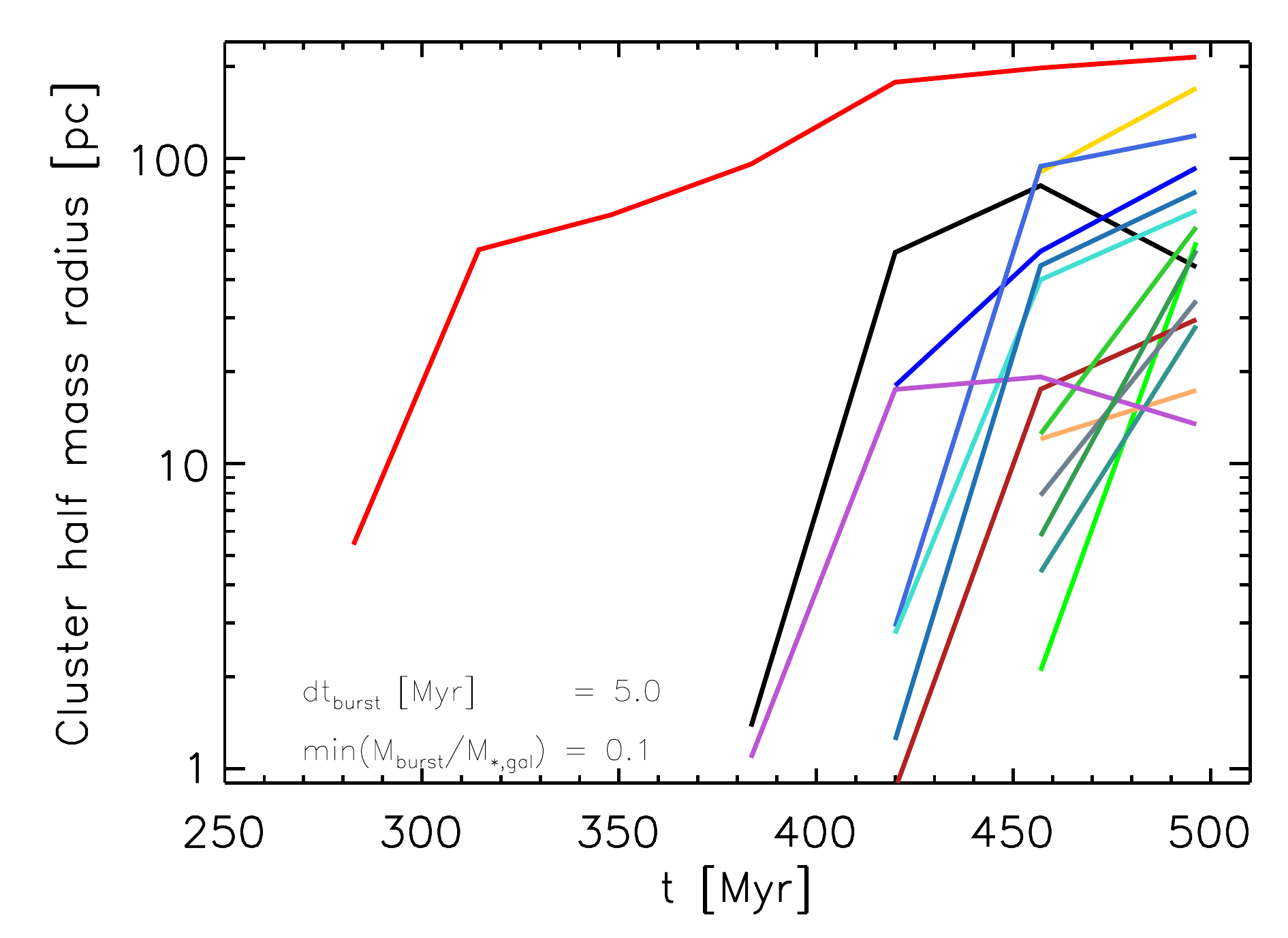}
\includegraphics[width=8.5cm]{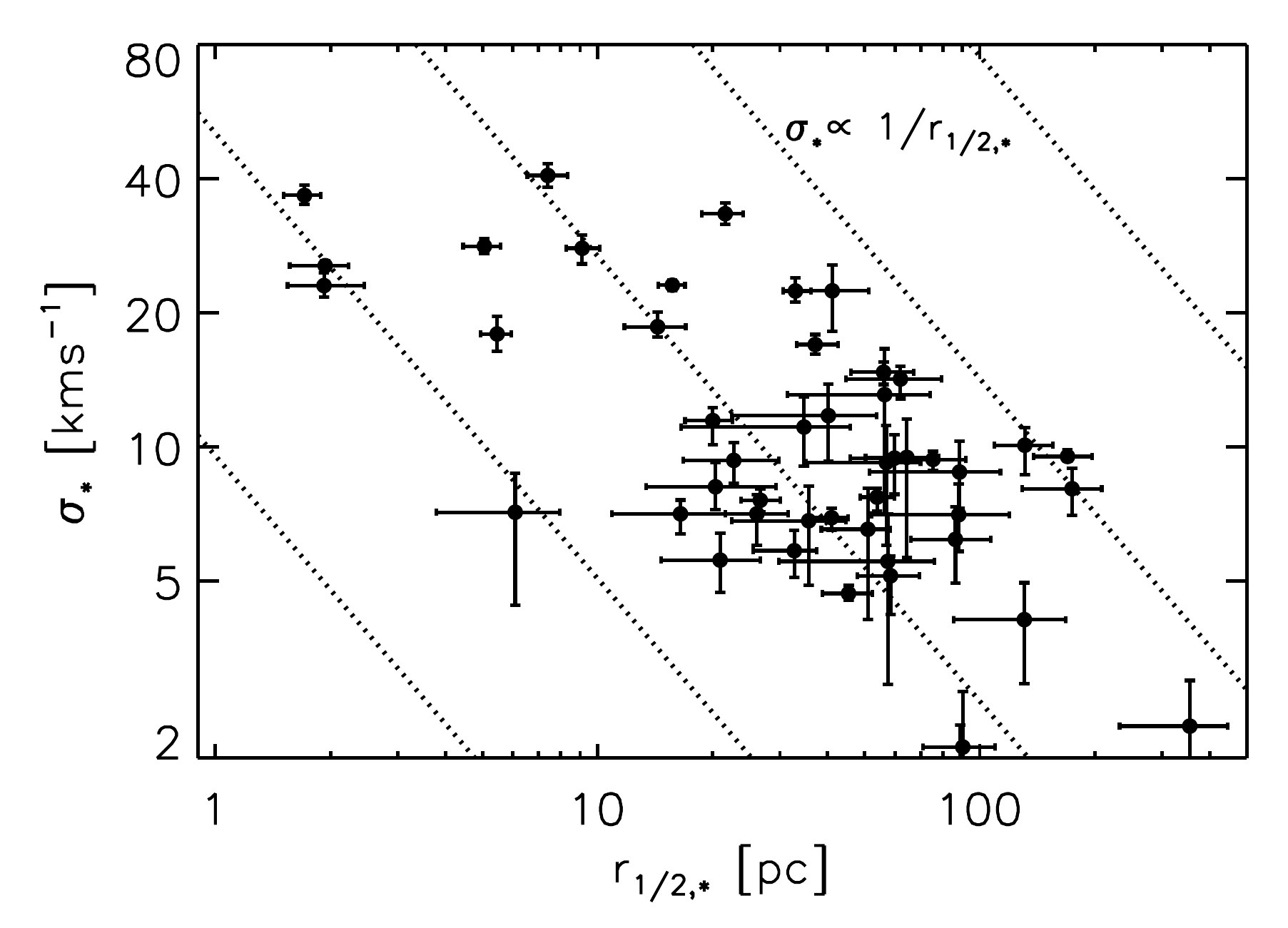}
\caption{(Left) Size evolution of star bursts (clusters) identified in
  the \RLREF simulation before the penultimate output (therefore
  tracks of compact clusters formed recently in the simulation do not
  appear here). (Right) 3D velocity dispersion of the stars as a function
  of the stellar half-mass radius for all bound objects with more than
  five star particles in the \RLREF simulation at $z=9$.}
\label{fig:cluster_evo}
\end{center}
\end{figure*}

In the left panel of \fig{fig:cluster_evo} we show how stars forming
in bursts tend to expand into more diffuse distributions over time.
``Clusters" are identified by dividing each galaxy's stars into 5~Myr
bins in formation time and then selecting any bins which account for
at least 10 per cent of the total stellar mass.  While there is no
explicit requirement that the stars in each burst be close to each
other, the majority have initial half-mass radii $<5$~\pc.  Each
selection is traced forward to subsequent output snapshots and its
half-mass radius is recorded.  In order to be plotted in
\fig{fig:cluster_evo}, bursts must occur before the penultimate
output.  We note that a number of the tracks show expansion from $\sim
5$~\pc to $\sim 50$~pc over a period of 100~Myr or so, while there is
tentative evidence that a few (between 1 to 3) remain fairly compact
($<30$\pc).  In many of the late-forming clusters, it is unclear
whether their final sizes will remain close to GCs radii or expand to
become similar to UFDs.

We can use a toy model to interpret the data from the simulation.  The
initial velocity dispersion of the stars in a proto-cluster,
$\sigma_*^{ic}$, can be estimated from the gas mass in the star forming
region of size $r_h^{ic} \sim 2$~pc:
\begin{equation}
\sigma_{*}^{ic} \sim \left(\frac{r_h^{ic}}{t_{dyn}}\right) \sim 50~{\rm km~s}^{-1} \left(\frac{r_h^{ic}}{2~pc}\right),
\end{equation}
where $t_{dyn}=(G\rho_{gas})^{-1/2} \sim 0.1$~Myr for
$\rho_{gas}=10^{-18}$\gpcc, which is appropriate at the maximum level
of refinement in our simulation (see \fig{fig:gas_images}).  Thus, the
initial velocity dispersion of the proto-star clusters is
$\sigma_*^{ic} > v_{\rm cir} \sim 10-20$\kms, where $v_{\rm cir}$ is
the circular velocity of the dark matter halo.  However, if the star
cluster is self-gravitating and bound, the stars will not be able to
escape the potential of the dark matter halo. Instead, if the star
cluster becomes unbound as a result of gas mass loss, its radius will
increase and the velocity dispersion of the stars will decrease.

Next we consider the effect of mass loss on the dynamical evolution of
a stellar system \citep{Hills1980}.  If the initial mass of the star
forming cloud is $M_{gas}^{ic}$ and the final mass after star
formation and gas loss is $M_*$, we can define the star formation
efficiency in the proto star cluster:
$\epsilon_{cl}=M_*/M_{gas}^{ic}$. There are two limiting cases.

\noindent
If $t_{\rm loss}\ll t_{\rm dyn}$ (impulsive gas loss):\\
\begin{align}
\frac{r_h}{r_h^{ic}} &= \frac{\epsilon_{cl}}{2\epsilon_{cl}-1}~~\text{with $0.5<\epsilon_{cl}<1$},\\
\frac{\sigma_*}{\sigma_*^{ic}} &\approx \left(\epsilon_{cl}\frac{r_h^{ic}}{r_h}\right)^{1/2}.
\label{eq:impuls}
\end{align}
In this case, only if $\epsilon_{cl}>50$\% will the cluster remain
bound.  The velocity dispersion of the stars decreases as $\sigma_*
\propto r_h^{-1/2}$ as the cluster expands to the new virial equilibrium
after mass loss\footnote{To derive the velocity dispersion we have
  applied the virial theorem to the final bound configuration, but the
  equation is nearly identical to Eq.~26 in \citet{Hills1980} for the
  expansion velocity of unbound associations.} (for
$\epsilon_{cl}=50\%$, $r_h\rightarrow \infty$ and $\sigma_* \rightarrow
0$).

\noindent
If $t_{\rm loss}\gg t_{\rm dyn}$ (quasi-adiabatic expansion):\\
\begin{align}
\frac{r_h}{r_h^{ic}} &= \frac{1}{\epsilon_{gc}}~~\text{with $0<\epsilon_{cl}<1$}, \\
\frac{\sigma_*}{\sigma_*^{ic}} &\approx \frac{r_h^{ic}}{r_h}.\label{eq:adiab}
\end{align}
In the right panel of \fig{fig:cluster_evo} we plot the velocity
dispersion of stars, $\sigma_*$, as a function of the half-mass
radius, \rhalf, for the galaxies in the \RLREF simulation. We observe
a bimodal distribution of $\sigma_*$: several galaxies
(about 12) are found to have $\sigma_*\sim20-40$\kms and
\rhalf$\sim1-20$\pc, while the rest are concentrated in the parameter
space $\sigma_* \sim 10 \pm 5$~km/s and \rhalf$\sim100\pm80$\pc. In
the plot we also show lines with $\sigma_* \propto \rhalfeqn^{-1}$,
consistent with quasi-adiabatic expansion of the cluster, as given by
\eqn{eq:adiab}.

Thus, combining the results illustrated in both panels of
\fig{fig:cluster_evo}, a picture emerges in which the low-surface
brightness dwarfs with \rhalf$\sim100$~\pc and $\sigma_*\sim10$~\kms
are the (young) descendants of dwarfs galaxies that form their stars
in compact clusters with high stellar velocity dispersions. A fraction
of these clusters with the highest star formation efficiencies remain
bound and resemble today's GCs, ultra-compact dwarfs or dwarf-globular
transition objects, while the others expand in the dark matter halo
potential until the stellar velocity dispersion (that decreases as
$\sigma_* \propto r_h^{-\alpha}$, with $\alpha \sim 0.5-1$) becomes
comparable to the halo circular velocity $v_{\rm cir}$ at the radius $r_h$:
\begin{equation}
\sigma_*(r_h) = v_{\rm cir}(r_h).
\label{eq:equil}
\end{equation}
At this point, the cluster is dark matter dominated and bound by the
gravitational potential of the dark matter halo. Thus, in this model a range of
stellar half-mass radii are possible, depending on the initial
$\sigma_*^{ic}$ and efficiency of star formation in the cluster
$\epsilon_{cl}$. However, $r_h$ cannot exceed $r_{\rm max}$ of the
halo (where the circular velocity reaches its maximum value). If this
happens, most of the stars will be lost from the dwarf into the IGM.
 
Assuming the cluster becomes unbound and evolves quasi-adiabatically,
and integrating $dr_h/dt=\sigma_*(r_h)$, obtained from
dimensional analysis, we get
\begin{equation}
\frac{r_h(t)}{r_h^{ic}} = \left(\frac{t}{t_{\rm dyn}}\right)^{1/2}
\label{eq:tracks}
\end{equation}
where $t_{\rm dyn} \equiv r_h^{ic}/\sigma_*^{ic}\sim
0.1$\Myr. Comparing $r_h(t)$ in \eqn{eq:tracks} as a function of time
with the evolutionary tracks in the left panel of
\fig{fig:cluster_evo} we find good agreement between our toy model and
the simulated clusters.

\subsection{Comparison to Present-day Compact Clusters and Nearby Dwarf Galaxies}
\begin{figure*}[th!]
\begin{center}
\includegraphics[width=18cm]{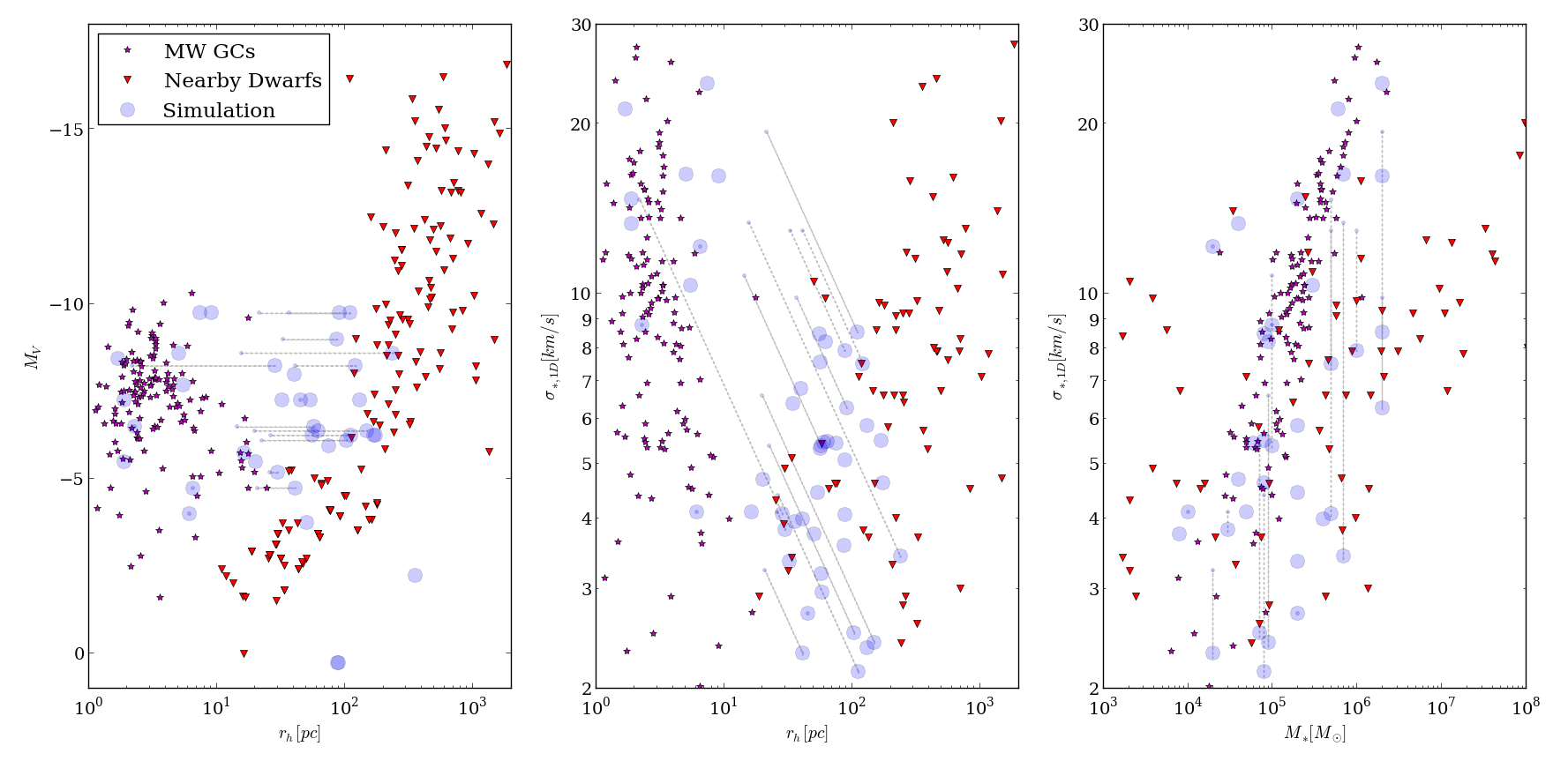}
\caption{Comparison between simulated compact clusters and dwarf
  galaxies after a simple extrapolation of stellar properties from
  $z=9$ to $z=0$ (blue circles), to Milky Way globular clusters
  (stars) and nearby classical and ultra-faint dwarf galaxies
  (triangles). We assumed a mass to light $M/L=3$. The dotted
  segments show the expected evolution of clusters that are either
  unbound or still have $>50\%$ of the their mass is in gas (see
  text).  ({\it Left.}) Visual absolute magnitude as a function of
  half-light radius, $r_h$. ({\it Center.}) Line of sight velocity
  dispersion as a function of $r_h$. ({\it Right.}) Line of sight
  velocity dispersion as a function of stellar mass.}\label{fig:obs}
\end{center}
\end{figure*}

In Figure~\ref{fig:obs} we show a comparison between the properties of
luminous objects in our \RLREF simulation (blue circles) in comparison
to Milky Way globular clusters (shown as stars) from \cite{Harris1996} and
nearby dwarf galaxies (shown as triangles), including classical dwarfs
(dSphs, dEs, dIrrs) and ultra-faint dwarfs from the
\cite{Mcconnachie2012} compilation. Similar plots for a compilation
of GCs and compact dwarfs are shown in \citet{Kissler-Patig2006}. Since our
simulation stops and is analyzed at redshift $z=9$, in order to
compare the simulated galaxies to present-day clusters and dwarf
galaxies, we assume a mass to light ratio $M/L_V=3$, that takes
into account mass loss and passive stellar evolution over about
12~Gyrs as in \cite{RicottiGnedin2005}. Several compact clusters in our
simulation are recently formed, thus they may not remain bound if i)
they are not in virial equilibrium, or ii) they contain a significant
amount of gas ($>50\%$) within their half-light radius, that if
expelled by SNe of photoevaporation may unbind the cluster (only a
couple of objects are in this category).  For those objects we evolve
their half light radii $r_h(t)$ as in Eq.~(\ref{eq:tracks}) and their velocity
dispersions $\sigma_*(t)$ as in Eq.~(\ref{eq:adiab}) until the stars becomes bound
by the dark matter halo as in Eq.~(\ref{eq:equil}). The dotted lines in the plot show the
evolution of such objects, evolving from the small to the larger blue
circles at the extremes of the dotted lines.  Note that here we plot
the line of sight velocity dispersion, $\sigma_{*,1D}$, that we simply
relate to the 3D velocity dispersion as
$\sigma_*=\sqrt{3}\sigma_{*,1D}$. Also, in order to calculate the
velocity profile $v_{\rm cir}(r)$ of each halo we assume NFW density
profile with concentration parameter $c=4$, that is appropriate for
recently virialized halos. We see that many simulated objects that are
initially in a region of parameter space devoid of observed objects (a
narrow strip laying between GCs and UFDs), are evolving to lower
$\sigma_{*,1D}$ and larger $r_h$ toward the region occupied by UFDs.


The extrapolation of observed properties to $z=0$ is rather
simplistic, and is not the main focus of this paper. Using N-body
simulations, previous works by \cite{BovillRicotti2011a,
  BovillRicotti2011b} have looked in detail at several effects that we
have neglected. They show that only small mass halos that are rather
isolated at $z=9$ evolve to $z=0$ without merging into larger halos
and thus accreted fresh gas. A subset of these halos evolving in
isolation with maximum circular velocity $v_{max} < 20-25$~km/s will
not accrete fresh gas from the IGM after reionization at $z \simlt
6-9$. These object, are suitable for a direct comparison to UFDs and
are what has been defined ``fossil'' galaxies
\citep{RicottiGnedin2005, BovillRicotti2009}. Bound compact cluster
may also survive intact for about a Hubble time (13.6 Gyrs) only if
they are more massive than $10^4$~M$_\odot$
\citep[\eg,][]{KatzR2014}. If their mass is instead smaller, the time
scale for evaporation due to 2-body encounters is shorter than the
Hubble. This is interesting because the faintest UFDs have $M_* \sim
10^3-10^4$~M$_\odot$, thus they could be formed by the secular
evaporation of compact clusters residing in a small mass dark matter
halo. As discussed above, the cluster will puff up until the stars
become bound by the potential of the dark matter halo, as in
Eq.~(\ref{eq:equil}). Of course GCs can be destroyed by tidal stripping
and shocks while interacting with their host galaxy . Roughly, only 1
or 2 in 10 GCs is expected to survive within the Milky Way over a
Hubble time \citep{KatzR2014}.
 
\begin{figure}[th!]
\begin{center}
\includegraphics[width=8.5cm]{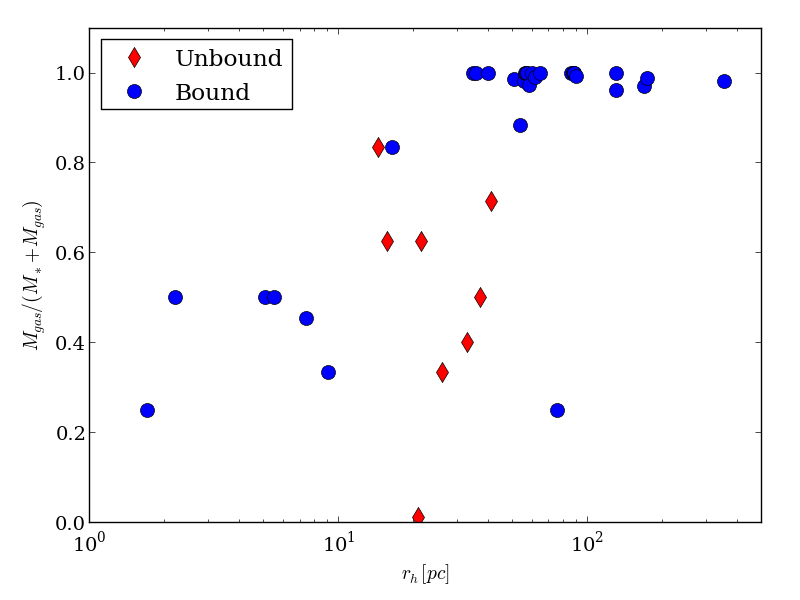}
\caption{Gas mass within the half-light radius, $M_{\rm gas}$,
  normalized by the total baryonic mass $M_* + M_{\rm gas}$ as a
  function of $r_h$.The circles show bound object and the diamonds
  unbound objects.}\label{fig:mgas}
\end{center}
\end{figure}
Fig~\ref{fig:mgas} shows the mass in gas within the half-light
radius, $M_{\rm gas}$, normalized by the total baryonic mass $M_* +
M_{\rm gas}$ as a function of $r_h$. In compact clusters with
$r_h<10$~pc the mass in gas is less than the mass in stars, while in
larger objects the mass of the stars is sub dominant with respect to
the gas mass (and the dark matter mass). The points shown as diamonds
are unbound objects, thus are transitioning toward larger $r_h$, while
the circles show bound objects (bound by the baryons if $r_h<10$~pc
and by the dark matter for $r_h>30$~pc). The figure shows that in
compact clusters either $>50$\% of the gas has been used for star
formation (\ie, the star formation efficiency in the proto-cluster is $\epsilon_{cl}
>50$\%), or $\epsilon_{cl}<50$\% and part of the gas has been
expelled by radiation and SN feedback. In the former case
($\epsilon_{cl}>50$\%) the cluster remains bound while in the
second case it will become unbound once all the gas is lost, and it
will expand untill $\sigma(r)=v_{cir}(r)$ (see \S~\ref{sec:disc}). The
figure shows that radiation and SN feedback did not clear out all the
gas withing the dark matter halo of most objects with $r_h>30$~pc, as
they are gas rich. However, the star formation rate in galaxies is
self-regulated by feedback (see \S~\ref{sec:subgrid} and Fig.~\ref{fig:rh_sfe}) and the
range of $\epsilon_{cl}$ found in the simulation is determined by the effectivness of
feedback in terminating star formation in star forming gas clumps and
by their density (the sub-grid star formation law we use converts into stars
$\epsilon_*=10$\% of the gas in a cell per local free-fall time).
We should however note that with a spatial resolution of 1~pc, the
internal structure of young compact star clusters ($r_h \sim$ few -
10~pc) is only marginally resolved. Higher resolution simulations are
needed to confirm quantitative results in the simulation. For instance
the number of bound stellar clusters may increase by increasing the
numerical resolution.

Multiple stellar populations are present in all globular clusters
observed to date \citep[\eg,][]{Grattonetal2004,DErcoleetal2008,Carrettaetal2010}.
\fig{fig:agemet_halos} shows that when we look in detail at the stellar populations of
compact star clusters, several have stars with distinct metallicities
and formation times. Some compact clusters form in very low mass
statellites orbiting the host dwarf galaxy (offset from center of
their host), triggered by external metal enrichment from galactic
winds. Others are found close to the center of their host dark matter
halos, resambling in terms of their metallicity the nuclei of compact
dEs. These second type of compact clusters should be able to form
stars in multiple bursts still remaining compact, due to gas fallback
in the gravitational potential of the dark matter halo. We do not have
the necessary resolution to study in detail the process of SN feedback
and gas fall back, but is an idea worth exploring with dedicated
high-resolution simulations.

\section{Summary and Conclusions}\label{sec:conc}

We have presented results from cosmological simulations of the
formation of the first stars and galaxies performed with an adaptive
mesh hydrodynamics code.  The code includes recipes for \popIII and
\popII star formation, self-consistent radiative transfer and a model
for the formation and dissociation of molecular hydrogen, allowing us
to resolve internal structure within the molecular clouds of
individual galaxies.

We find that the first galaxies have thick gaseous disks but the stars
form spheroids with half light radii larger than the disk
thickness. Within the disk, star formation takes place in compact star
clusters with initial velocity dispersions $\sigma_* \sim 20-40$~km/s
and half-mass radii of a few parsecs. However, due to gas loss, most
of the clusters become unbound and expand, until the stars become
bound again by the gravitational potential of the dark matter
halo. This is confirmed by our analysis of the simulations showing
that the stars in low surface brightness dwarf galaxies were in much
denser star clusters at the time of their formation.  These clusters
had sufficiently high initial velocity dispersions to expand to
$\sim100$\pc in a few Myrs after gas loss following SN and radiation
feedback.  Thus, the main reason for the formation of stellar
spheroids with little rotation in the first dwarfs appears to be the
high velocity dispersion of stars in the proto star clusters, relative
to the circular velocity of the host halo ($10-20$\kms).  There is
some tentative evidence that we have captured the formation of the
first bound compact stellar systems (although is unclear their
classification as proto-GCs, ultra-compact dwarfs or dwarf-globular
transition objects), as about 1 to 3 of the clusters with initially
compact configuration remain bound for a few hundred Myrs, until the
end of the simulation at $z\sim 9$. Higher resolution
simulations evolved to lower redshift are necessary to confirm this
result. However, the origin of low surface brightness dwarfs as evaporated
compact stellar clusters appears robust, suggesting a new connection
between UFDs (or dSphs) and compact stellar system. We therefore speculate that:
\begin{enumerate}
\item A fraction of old GCs around the Milky Way, although clearly not
  all of those observed, may form within dark matter satellites around low
  mass halos in the early universe (see triggered star formation in
  \S~\ref{sec:prop} and Figure~\ref{fig:Mdm_r}).
\item Observational signatures of a dark matter minihalo surrounding
  GCs (in addition to dynamical evidence for dark matter in the
  outer parts of the cluster) would include: an enhanced ability to
  retain metals during their formation and possible modifications to
  the effects of dynamical friction and tides on their secular
  evolution.
\item Candidate compact clusters are offset from the center of
  their host halos by an average of $\sim 40$\pc.  Some are found in
  \appmass{8} halos that contain both \popIII and \popII stars, but a
  few are found in very low mass (\appmass{6}) halos that do not
  contain \popIII stars. This indicates that the formation of bound
  stellar clusters was triggered by the influence of a nearby luminous
  galaxy. Their formation in \appmass{6} satellites of \appmass{8}
  halos may help explain the spatial distribution of GCs in nearby
  dwarf galaxies.
\item In our simulations UFDs originate from the
  dissolution of only a few distinct star clusters. Open clusters, or
  compact clusters with masses $\simlt 10^4$~M$_\odot$ (that evaporate
  in a Hubble time due to 2-body encounters) will expand inside the
  dark matter halo, resulting in a low surface brightness, dark matter
  dominated galaxy. It may be possible to identify stars that
  originated in the distinct progenitor clusters through their
  chemical and kinematic properties.
\item The faintest dwarfs found within $150$~\kpc of the Milky Way
  have half-mass radii of $20-40$~\pc and hence are typically less
  extended, with higher surface brightnesses than some brighter
  satellites.  In our simulations such dwarf galaxies, with properties
  intermediate between \rhalf$\sim100$~\pc UFDs and GCs exist. The
  stellar half light radii of these objects are set by the initial
  velocity dispersion of the stars in the proto-cluster and the gas
  mass loss rate. However, it is less clear if the fossil remnants of
  these compact objects will remain stable to the present (for
  $12$~Gyr), or if their stellar component will expand further during
  their secular evolution and interaction with Milky Way.
\item The metallicity and duration of star formation in some compact
  stellar clusters are consistent with a single star burst.  However,
  several compact dwarfs with sizes $\sim 5-40$~\pc have one or more
  additional subdominant stellar components.  These superimposed
  components are typical of dwarf galaxy formation in the
  early universe as illustrated in \fig{fig:agemet_halos}: a rather
  bursty and spatially segregated star formation history spread over a
  few 100~\Myr, and a metallicity distribution produced by
  the hierarchical assembly of several smaller dwarfs
  \citep{RicottiGnedin2005}.
\end{enumerate}
The simulations presented here and in PRG16 are only a first step to
understand the emergence of the first light in the universe. The
qualitative results are intriguing because they suggest interesting
ideas on the origin of compact star clusters and ultra-faint dwarfs
and a deeper physical understanding of their connections. However, in
order to answer questions on the statistics of the relics of the first
objects in the Local Group, a larger computational volume and higher
dark matter resolution will be necessary to achieve numerical
convergence.

\section*{Acknowledgments}
OHP and MR acknowledge support from NASA grant NNX10AH10G and NSF
grant CMMI1125285.  The authors acknowledge the University of Maryland
supercomputing resources (http://www.it.umd.edu/hpcc) made available
in conducting the research reported in this paper.  This work also
used the Extreme Science and Engineering Discovery Environment
(XSEDE), which is supported by NSF grant number ACI-1053575.

\bibliography{bibliography1}

\end{document}

%% file: useful_macros.tex
\usepackage{xspace}

\def\gtsim{>\kern-1.2em\lower1.1ex\hbox{$\sim$}~}  
\def\simgt{\lower.5ex\hbox{\gtsima}}
\def\ltsim{<\kern-1.2em\lower1.1ex\hbox{$\sim$}~}  
\def\simlt{\lower.5ex\hbox{\ltsima}}

\def\ltsima{$\; \buildrel < \over \sim \;$}
\def\gtsima{$\; \buildrel > \over \sim \;$}

\newcommand{\eqn}[1]{Equation~(\ref{#1})}
\newcommand{\fig}[1]{Fig.~\ref{#1}}
\newcommand{\tab}[1]{Table~\ref{#1}}
\newcommand{\sect}[1]{Section~\ref{#1}}

\def\art{{\rmfamily\scshape art}\xspace}

\def\subfind{\textsc{subfind}\xspace}

\def\ie{{\it i.e.}}
\def\eg{{\it e.g.}}

\def\hmol{H$_2$\xspace}

\def\HI{{\ion{H}{i} }}

\def\popII{Pop~II\xspace}
\def\popIII{Pop~III\xspace}

\def\HI{\hbox{H~$\scriptstyle\rm I\ $}}

\def\HeI{\hbox{He~$\scriptstyle\rm I\ $}}
\def\HeII{\hbox{He~$\scriptstyle\rm II\ $}}

\def\lcdm{$\Lambda$CDM\xspace}

\def\rhalf{$r_{\rm h}$\xspace}
\def\rhalfeqn{r_{\rm h}}
\def\fhmol{$f_{\rm H_2}$\xspace}
\def\zcrit{$z_{\rm crit}$\xspace}

\newcommand{\appmass}[1]{\mbox{$10^{#1}~\rm{M_{\odot}}$}}
\newcommand{\solmass}[2]{\mbox{$#1 \times 10^{#2}~\rm{M_{\odot}}$}}

\def\hinvmpc{\xspace h$^{-1}$~Mpc\xspace}

\def\kms{\xspace{$\rm\,km\,s^{-1}$}\xspace}

\def\kpc{\xspace kpc\xspace}
\def\pc{\xspace pc\xspace}

\def\Myr{\xspace Myr\xspace}
\def\gpcc{\xspace${\rm gcm}^{-3}$\xspace}

\def\hide#1{}

\usepackage{xparse}